\documentclass[12pt]{article}

\usepackage{exscale}
\usepackage{amsmath, amssymb, amsthm}
\usepackage{feynmf}

\newcommand {\slashi}[1] {\rlap{\sl/}#1} 
\newcommand\slashii[1]{\setbox0=\hbox{$#1$}             
   \dimen0=\wd0                                 
   \setbox1=\hbox{\sl/} \dimen1=\wd1            
   \ifdim\dimen0>\dimen1                        
      \rlap{\hbox to \dimen0{\hfil\sl/\hfil}}   
      #1                                        
   \else                                        
      \rlap{\hbox to \dimen1{\hfil$#1$\hfil}}   
      \hbox{\sl/}                               
   \fi}                                         %
\newcommand\slashiii[1]{\setbox0=\hbox{$#1$}#1\hskip-\wd0\hbox to\wd0{\hss\sl/\/\hss}}
\newcommand\slashiv[1]{#1\llap{\sl/}}
\newcommand {\half} {{1 \over 2}}

\newcommand {\bra}[1] {\left< #1 \right|}
\newcommand {\ket}[1] {\left| #1 \right>}

\newcommand {\braket}[2] {\left< #1 \mskip1mu\vrule\mskip1mu #2 \right>}

\newcommand {\diag} {\mathrm{diag}\,}

\theoremstyle{remark}

\theoremstyle{definition}

\begin{fmffile}{fmfilezb}
\begin{document}

\title{
 Unitarity, Lorentz invariance and causality  in  Lee-Wick theories:\\
    An asymptotically safe 
    completion of QED
      }
    \author{Andr\'e van Tonder
            \\ \\
            Department of Physics, Brown University \\
            Box 1843, Providence, RI 02912, USA
            }
    \date{October  1, 2008}

    \maketitle

    \begin{abstract}
        \noindent
We revisit  the previously unsolved
problems of  ensuring Lorentz  invariance and non-perturbative
unitarity in Lee-Wick theories.  We base our  discussion
on an ultraviolet completion of QED by Lee-Wick ghost fields,
which  is argued to be asymptotically safe.
We argue that as long as the state space is based upon 
a suitable choice of distributions of a type invented 
by Gel'fand and Shilov, the Lee-Wick 
ghosts can be eliminated while preserving 
Lorentz invariance to produce a unitary 
theory.  The method for eliminating ghosts
is in principle non-perturbatively well-defined, in contrast with some previous 
proposals.  
We also point out a second, independent  mechanism for
producing a unitary theory, based on a covariant 
constraint  on the maximum four-momentum, which 
would imply an amusing
connection, based on naturalness,  between the coupling constant and the 
hierarchy of scales in the theory.
We further emphasize that the resulting 
theory is causal, and point out some analogies  between
between the behaviour of Lee-Wick ghost degrees of freedom and 
black holes.

          
        %
    \end{abstract}

\section{Introduction}

We argue that there is
justification for reconsidering  quantum field theories
of  Lee-Wick type as potentially realistic  fundamental or effective 
theories of  the world.  The original justification for these theories,
which remains as valid as ever,  
was their good ultraviolet behaviour.  
We believe that the methods 
expounded in this paper can be applied to solve  the 
problems related to unitarity, non-perturbative definition, 
 and 
Lorentz invariance identified by previous authors.

In the sixties, T.D.\ Lee and 
G.C.\ Wick initiated a program to  extract  ultraviolet complete unitary 
quantum field theories embedded in certain larger 
indefinite inner product field
theories \cite{leewick, lee1, lee2}.  These authors proposed 
extending certain quantum field theories by adding 
 ghost degrees 
of freedom, similar to the familiar Pauli-Villars
regulator fields \cite{PV},
to  make the theory finite by
 cancelling divergences.
Unlike Pauli-Villars ghosts, however, the Lee-Wick 
ghost masses were to be kept finite.  Lee and Wick argued that,
as long as all ghost degrees of freedom 
in the interacting theory 
could be shown to have complex energies, one could 
obtain a unitary theory by constraining the a physical 
subspace to be exactly those states that  have real energy.

However, this program ran into two serious obstacles that
were never satisfactorily resolved.
We propose to resolve both of these problems in the 
present article.  

First, it was quickly 
realized that in quantum field theories with the required 
complex mass ghosts, one would expect multi-ghost 
states that have real energy \cite{lee2}.  It seemed that 
these states could not be 
eliminated by the Lee-Wick real energy constraint,
and other
arguments would be needed to get rid of them.  
Ad hoc prescriptions to do so order by order
in perturbation theory were proposed by Lee \cite{lee2}, as well as 
by Cutkosky et al.\  \cite{cutkosky}, but no unambiguous all-order prescription
was ever found and, as  was argued among others by Boulware and Gross 
\cite{BG} and by
Nakanishi \cite{nakanishi, nakanishiprime}, it is  questionable
whether these prescriptions have any non-perturbative meaning.

Second, as was shown by Nakanishi and Gleeson  et al. 
\cite{nakanishi, gleeson}, 
a Hamiltonian-based non-perturbatively well-defined approach  to 
eliminating multi-ghost states seemed doomed to 
fail if we insist on Lorentz-invariance.  In particular, there 
did not seem to be a Lorentz-invariant way of 
applying the real-energy constraint in a Hamiltonian
formulation of  the theory.  

We will show first  
that the problem with Lorentz invariance 
can be solved in a non-perturbative setting by 
carefully constructing the multi-ghost state spaces 
so that they are Lorentz-invariant.  The problem with 
Lorentz invariance identified by Nakanishi and 
Gleeson  et al.\  can be traced to their use
of state spaces consisting of degrees of freedom with complex 
energies but real momenta.  Such a description is 
obviously not Lorentz invariant.  A Lorentz-invariant
construction has to address the problem of complex momenta.
Fortunately, a framework that can be used for such states 
exists in the form of a class  of spaces of generalized 
functions based on analytic  test functions that was 
invented by Gel'fand and Shilov \cite{gelfand1, gelfand2}. 
 The construction
of the required states is quite subtle, but we argue that 
it resolves this part of the problem.    

The resulting sets of multi-ghost states will be Lorentz-invariant.
However, at first glance, their energies would seem to be real, 
so that naively they would not seem to be eliminated by the
Lee-Wick real energy constraint.  However, we will show that once we take into account 
interactions in a careful non-perturbatively well-defined approach, these energies
will generally become complex, and as long as we apply the 
Lee-Wick real energy constraint and the infinite-volume limit in the 
correct order, these states will disappear from the 
physical spectrum, and the resulting theory can be 
expected to be unitary.

We illustrate these methods with
a simple extension of QED  with high-mass ghost particles
that we argue to be \textit{asymptotically 
safe} \cite{weinberg}.  By this we mean that the fine structure constant
remains small at all energies and approaches a fixed 
point at infinite energy.  Our proposal differs from Lee's in that 
the theory is only finite after  mass renormalizations.
Still, unlike ordinary QED, which 
is not believed to exist up to arbitrarily high energies,
our asymptotically safe extension should be 
 an ultraviolet complete theory.  
In particular, in a non-perturbative construction relying on 
some dicretization, 
asymptotic safety implies that the theory 
should have a  continuum limit, which  is a a prerequisite 
for exact Lorentz invariance.
We argue that all ghosts can be eliminated in a 
Lorentz-invariant  and non-perturbatively well-defined
way.

We also discuss a second (poor man's) method for getting rid
of ghost states.  A universe with finite mass-energy
could perfectly well be described by an ultraviolet-complete
theory containing ghosts, as long as the mass-energy of the 
universe is smaller than that of any real-energy ghost states.
In other words, as long as the ghost masses are 
large enough, the subspace of the state space 
available to the universe, or
any subsystem of the universe, is positive-definite, and the 
description of the world will be unitary.  Taking this 
idea a little further, we point out an interesting connection,
based on a naturalness  argument, between the large mass-energy
of the universe and the smallness of the fine structure
constant.

We also discuss the somewhat misleadingly named ``acausality'' of these
theories, and emphasize, as did previous authors 
\cite{lee2, BG, coleman, causality},
 that 
no inconsistencies can arise.  
We discuss, by way of example, how
a particular form of the grandfather paradox is avoided.
We propose that it may be 
less confusing to think of these theories as non-local, rather than
acausal, at 
sufficiently high energy scales.  
Initial states 
typically contain precursors (as they were called by Lee \cite{lee2})
 beyond the limits of experimental precision,  
that get 
exponentially 
magnified by the time evolution and  may only become observable
at later times.  For the reader uncomfortable with this aspect 
of these theories, we point out that these  precursors are 
morally not so different from the precursors invoked 
to explain aspects of the bulk-boundary correspondence 
in String Theory holography \cite{precursors1, precursors2}.   
We further point out that 
Horowitz and  Maldacena recently proposed resolving
the black hole information paradox via a final state 
boundary condition \cite{maldacena}. 
 Again, this is not so alien
in the context of Lee-Wick theories, since 
 the Lee-Wick real energy constraint 
can be reinterpreted as incorporating 
a final state boundary condition, 
namely, no blow-up at future infinity \cite{BG}.
In the work of Horowitz and  Maldacena, the
 black hole
final 
state boundary condition ensured that everything 
falling into the black hole annihilated completely, 
leaving nothing behind.  But as we shall see,
this is  analogous to what 
happens in Lee-Wick theories in typical 
scattering processes mediated by ghosts, 
where the incoming particles annihilate 
to form a null state that decays exponentially 
in the future.

It is worth noting that, while  our approach is  an adaptation 
of Lee and Wick's original proposal for eliminating ghost states,
recent work of Bender et al. \cite{bender1}
propose an alternative interpretation of these theories 
based on PT-symmetry.   Their approach differs from ours, which 
does not rely on PT-symmetry.

Given our proposed resolutions to some prior 
foundational difficulties 
encountered in Lee-Wick theories, 
we believe that these theories deserve another look
in the search for descriptions of
nature.  Indeed, Lee-Wick theories  have recently enjoyed
a revival in the form of the 
Lee-Wick Standard Model
\cite{grinstein, espinosa, rizzo, dulaney, alvarez, carone, underwood, causality},
proposed to address, among other things,  the hierarchy problem and 
the stability  of the Higgs mass. 
Hopefully the results of this paper can be generalized to 
provide solutions to the problems of non-perturbative unitarity 
and Lorentz invariance in these theories.  

In the light of our results,  it may also be productive to revisit 
renormalizable
higher-derivative modifications of gravity, which 
contain Lee-Wick degrees of freedom
\cite{stelle}.

\section{Ghost quantum electrodynamics}

We will show that quantum electrodynamics may be completed by
the addition of a ghost field to obtain a theory that will be argued to 
be asymptotically safe.  The construction is in principle 
non-perturbatively well-defined, and will be argued in subsequent sections to
satisfy the properties of Lorentz covariance and 
unitarity
necessary for a consistent 
physical interpretation, provided certain constraints are imposed on the 
state space to effectively eliminate the ghosts.  

Consider the Lagrangian
\begin{align}
  \mathcal L
&= \sum_{i=0}^1 \bar \psi_i \left(i \,(\slashi\partial - i\slashii A) - {m_i}\right) \psi_i
      - {1\over 4e^2} F_{\mu \nu} F^{\mu\nu}  
    -{\lambda\over 2e^2}\, \left(\partial_\mu A^\mu\right)^2 + \cdots
\nonumber \\ 
   &= 
   \sum_{i=0}^1 Z_{i\Lambda}\,\bar \psi_i \left(i \,(\slashi\partial - i\slashii A) - {m_i}_\Lambda\right) \psi_i
      - {1\over 4e_\Lambda^2} F_{\mu \nu} F^{\mu\nu}  
    -{\lambda_\Lambda\over 2e_\Lambda^2}\, \left(\partial_\mu A^\mu\right)^2.
  \label{action}
\end{align}
Here $\psi_0$ is an ordinary fermion, while $\psi_1$ is a 
Dirac boson  --
a particle with Dirac action that is quantized using bosonic
statistics \cite{indef}. 
 A theory containing such a field would ordinarily  violate unitarity, be unstable, or both.  
These problems will be avoided later by constraining the state 
space. 

The first line shows the  physical fields,
masses,  and renormalized charge, defined as usual
in terms of the positions and residues of poles of the full propagators.
The second line shows the full bare action, whose form is 
constrained by gauge invariance, in terms of the 
bare masses, field renormalizations 
 and charge at regularization
scale $\Lambda$.  
The dots on the first line stand for 
the counter-terms that make up the difference.
In electrodynamics, the gauge fixing term 
does not get renormalized, so $\lambda/e^2 = \lambda_\Lambda/e_\Lambda^2$.
 
In contrast to the
situation in QED,
we shall argue that, under certain conditions, 
 the theory is asymptotically safe and 
there is no 
obstruction to taking the limit
 $\Lambda \to \infty$.  

This type of action is not entirely new.  Bosonic fields of Dirac type
have been used for many years as Pauli-Villars regulator fields.  However,
whereas Pauli-Villars regulator  masses are taken to infinity as part of
renormalization, we keep all masses finite.  A similar, though not 
identical, proposal was made by Lee with the aim of
making QED finite \cite{lee1}.  To achieve finiteness, 
 Lee  also included a 
massive counterpart to the photon.  Since we 
do not include a massive vector particle, our theory will be finite only 
once we perform 
mass renormalizations.  In addition, 
whereas Lee's extra Dirac fields were 
given complex masses, all masses appearing in our
action will be assumed real.

\section{Dirac bosons}
  
Dirac bosons are discussed in \cite{indef}.
A Dirac boson can be quantized in a way that satisfies 
micro-causality (locality) using either a positive-definite or an indefinite 
state space inner product.  

Choosing the positive-definite representation would lead directly to an 
unstable theory in which the energy is unbounded below.  
We will therefore consider instead the indefinite representation, in which 
the unperturbed energies are nonnegative.
\begin{align*}
  \psi (x) &= \int {d^3 \mathbf p\over (2\pi)^3}\,{1\over \sqrt{2E_{\mathbf p}}}
   \sum_s \left( a_{\mathbf p}^s u^s (p)\,e^{-ip\cdot x}
   +  b_{\mathbf p}^{s\dagger} v^s (p)\,e^{ip\cdot x}
   \right), 
\end{align*}
where
\begin{align*}
   [a_\mathbf{p}^s, a_\mathbf{q}^{r\dagger}] &= (2\pi)^3\, \delta^3(\mathbf p -
      \mathbf q)\, \delta^{rs}, \\
[b_\mathbf{p}^s, b_\mathbf{q}^{r\dagger}] &= - (2\pi)^3\, \delta^3(\mathbf p -
      \mathbf q)\, \delta^{rs}, \\
  a_\mathbf{p}^s \ket {0} &= 0, \\
  b_\mathbf{p}^s \ket{0} &= 0.
\end{align*}
In this representation, any state containing an odd number of the bosonic
$b$-anti-particles  has 
negative metric.  The propagator is the same as for a 
Dirac fermion, and is given by
$$
  i\over {\slashiv p - m + i0}.
$$
However, because the field is bosonic, loops do not come with a 
factor $-1$.    
As a result, the presence of Dirac bosons in ghost QED will 
cause high-energy loop diagram cancellations that will
make the theory asymptotically safe.

Note that the $i0$-prescription does not follow from 
a conventional convergence factor argument.  It can, however, be derived from 
the operator representation and also from a properly defined non-perturbative 
path integral \cite{indef}.  

Since the action is real-valued, the associated
 Hamiltonian, which acts on an indefinite inner product space,
will be pseudo-hermitian.
Pseudo-hermitian operators may have, in addition to real eigenvalues,
complex eigenvalues occurring in conjugate pairs and associated to 
null eigenmodes.  
Since such modes grow or decay exponentially with time, 
their production would signal an instability 
in the indefinite inner product 
theory, and we will have to find a way of avoiding production of
such states.  

A more serious problem with indefinite representations is the 
issue of unitarity.  These representations 
contain states whose inner product with 
themselves is negative, also sometimes called negative-metric states. 
 In our case, any state containing an
odd number of Dirac boson anti-particles is a negative metric state.
The presence of these states would be a disaster for the 
probability interpretation unless there is a mechanism for 
eliminating these states or preventing their production.

We will later impose constraints on the state space 
under which the interacting theory remains stable and unitarity will be satisfied.

\section{Asymptotic safety}
\label{sectionsafe}

We first argue that the theory defined by (\ref{action}) is asymptotically
safe \cite{weinberg} 
for a large region of the parameter space.   In other words, the 
fine structure constant
$\alpha_\Lambda = e^2_\Lambda/4\pi$ will remain a finite small parameter,
approaching a fixed point, as $\Lambda \to \infty$.  We will
motivate this by a calculation to second order, and conclude with a general
argument that the theory should remain asymptotically safe to all orders.  

In contrast with the case of 
ordinary quantum electrodynamics, 
ghost loops will cancel the infinities that would
otherwise appear in the vacuum polarization, as a result of which
the bare charge does not need an infinite 
renormalization.  Therefore, provided the bare charge 
remains small under renormalization,
we are not required to separate an $F_{\mu\nu}F^{\mu\nu}$ counter-term
and are free to organize our perturbation
expansion in terms of the bare charge $e_\Lambda$, writing the action as
\begin{align}
  \mathcal L
&= \sum_{i=0}^1 \bar \psi_i \left(i \,(\slashi\partial - i\slashii A) - {m_i}\right) \psi_i
- {1\over 4e_\Lambda^2} F_{\mu \nu} F^{\mu\nu}  
    -{\lambda_\Lambda\over 2e_\Lambda^2}\, \left(\partial_\mu A^\mu\right)^2\\
   &\quad +
   \sum_{i=0}^1 (Z_{i\Lambda} - 1)\,
\bar \psi_i \left(i \,(\slashi\partial - i\slashii A) - {m_i}\right) \psi_i 
- Z_{i\Lambda} ({m_i}_\Lambda - m_i)\, \bar \psi_i \psi_i.
  \label{action1}
\end{align}
The counter-terms in the second line may be expressed as functions 
of $\Lambda$ and $e_\Lambda$, for which an asymptotic expansion 
in $e_\Lambda$ 
may be derived, as usual, order by order in perturbation theory.  
As we shall discuss, for a large range  of parameters,
no large logarithmic corrections will spoil perturbation theory, and
all mass terms will be real.  Mass renormalization will be 
discussed in the next section.

Consider the one-loop correction to the vacuum polarization
$\Pi_{\mu\nu} (q)$, given 
by the diagrams 
\newenvironment{Zff}
{\begin{fmfgraph*}(35,25)
    \fmfleft{G}\fmfright{GG}
  }
  {\end{fmfgraph*}}

\begin{equation}
\parbox{40mm}
{\begin{Zff}
  \fmf{boson}{G,v1}\fmf{boson}{v2,GG}
  \fmf{fermion,left,tension=.5}{v1,v2}
  \fmf{fermion,left,tension=.5}{v2,v1}
  \fmfdot{v1,v2}
\end{Zff}} 
+
\hskip 5mm 
\parbox{50mm}
{\begin{Zff}
  \fmf{boson}{G,v1}\fmf{boson}{v2,GG}
  \fmf{scalar,left,tension=.5}{v1,v2}
  \fmf{scalar,left,tension=.5}{v2,v1}
  \fmfdot{v1,v2}
\end{Zff}}
\end{equation}
that are superficially quadratically divergent.
In fact, the correction is finite, as a consequence of gauge invariance
and the ghost contribution.

Gauge invariance requires the current conservation Ward identity
\begin{align}
  q^\mu \Pi_{\mu\nu} (q) = 0, \label{ward}
\end{align}
which in turn implies that no quadratically divergent contribution 
can occur in a gauge-invariant regularization.\footnote{
  For example, consider the Pauli-Villars regularization 
obtained by adding an equal number of Dirac bosons and Dirac fermions
whose masses are to be taken to infinity.  
The result is
$$
  {e_\Lambda^2\over 2\pi^2}\, (q_\mu q_\nu - g_{\mu\nu} q^2)
   \int_0^1 d\beta \, \beta(1-\beta)\sum_i c_i\,\ln{\Lambda_0^2\over{m_i^2 - \beta(1-\beta)\,q^2}}-{e_\Lambda^2\over 8\pi^2}\, g_{\mu\nu}\sum_i c_i\, m_i^2,
$$
where the sums include physical, ghost and  regulator fields, and the
 $c_i=\pm 1$
denote the statistics.  
 The first term is independent of the cutoff
$\Lambda_0$ and can be written in the form (\ref{Pimunu0}), where $\Lambda$ is
a function of the regulator 
masses, for large values of the latter.
The second term is
only consistent with the Ward identity (\ref{ward}) if the 
Pauli-Villars mass condition $\sum_i c_i m_i^2 = 0$
is imposed.  Thus, gauge invariance prohibits the quadratic divergence.
}
The remaining logarithmic divergence
is canceled by the ghost contribution, as is clear from the
fact that
\begin{align}
  \Pi_{\mu\nu} (q) &=  (q_\mu q_\nu - g_{\mu\nu} q^2)\, \Pi (q^2),
\end{align}
where
\begin{align}
\Pi (q^2)
&=  {e_\Lambda^2\over 2\pi^2}
   \int_0^1 d\beta \, \beta(1-\beta)\sum_i c_i\,
\ln{\Lambda^2\over{m_i^2 - \beta(1-\beta)\,q^2 - i0}},
    \label{Pimunu0}
\end{align}
valid as long as  $|q^2| \ll \Lambda^2$.  
The logarithmic term
is independent of the regulator
scale $\Lambda$
due to our choice of statistics
$$
  \sum_i c_i = 0,
$$
where  $c_i = 1$ for a fermionic
field and $c_i = -1$ for a bosonic field.  
We obtain
\begin{align}
  \Pi (q^2) &=  {e_\Lambda^2\over 2\pi^2}
   \int_0^1 d\beta \, \beta(1-\beta)\,
\ln{{M^2 - \beta(1-\beta)\,q^2 - i0}\over{m^2 - \beta(1-\beta)\,q^2 - i0}},
\qquad |q^2| \ll \Lambda^2.
    \label{Pimunu1}
\end{align}
Here $m = m_0$ denotes the physical
lepton mass and  $M = m_1$ denotes the ghost 
mass.  

No infinite renormalization of $e_\Lambda$  is needed to make 
this well-defined as $\Lambda \to \infty$, and 
we therefore take the bare charge to be a finite 
value independent of $\Lambda$, i.e.,
$$
  e_\Lambda = \textrm{constant} \equiv  e_\infty,
$$
and take $\Lambda \to \infty$ to obtain
\begin{align}
  \Pi (q^2) &=  {e_\infty^2\over 2\pi^2}
   \int_0^1 d\beta \, \beta(1-\beta)\,
\ln{{M^2 - \beta(1-\beta)\,q^2 - i0}\over{m^2 - \beta(1-\beta)\,q^2 - i0}},
    \label{Pimunu}
\end{align}
now valid for all $q^2$.

The effective coupling $\alpha_\mu = e_\mu^2/4\pi$ at space-like momentum 
transfer 
$$
  q^2 = -\mu^2
$$
is then given by a geometric series over 1PI diagrams, which gives to this order,
\begin{align}
  \alpha_\mu &= {\alpha_\infty \over 1 + \Pi(-\mu^2)} \nonumber \\
  &= {\alpha_\infty \over \displaystyle{1 + {2\alpha_\infty\over\pi}
      \int_0^1 d\beta \, \beta(1-\beta)\,\ln {{M^2 + \beta(1-\beta)\,\mu^2 }\over{m^2 + \beta(1-\beta)\,\mu^2}}}}. \label{alphamu}
\end{align}
The geometric series converges as long as $\Pi(-\mu^2) < 1$, which will
be the case as long as the bare coupling is small enough  that
$$
  {\alpha_\infty \over 3\pi}\, \ln{M^2\over m^2} < 1
$$
since the logarithmic term is a monotonic decreasing function of 
$\mu$ if $M>m$.   We will discuss the physical implications of this bound 
below.  

As in ordinary QED, the effective coupling grows with increasing energy.
However, in contrast to ordinary QED, the effective coupling is well-defined
at all energy scales, and asymptotically 
approaches the finite bare coupling $\alpha_\infty$
at infinite energy or $\mu\to\infty$.\footnote{For time-like momentum transfer $q^2>m^2$ the 
vacuum amplitude has an imaginary part.  But this vanishes as
 $q^2 = -\mu^2\to \infty$, so in this limit we find the same
effective coupling $\alpha_\Lambda$.}  The effective coupling is always smaller
than the bare coupling, so if the bare coupling is chosen sufficiently
small, perturbation theory can be trusted for all scales. 
In other words, 
to this order in perturbation theory the theory appears to  be
asymptotically safe.

To this order, then, the 
theory will be free of the Landau pole, the unphysical tachyonic 
singularity at large
space-like momentum transfer (short space-like distances)
appearing in perturbative QED.  While such perturbative arguments
are not conclusive, the existence of 
this divergence in ordinary QED is usually regarded as strong 
evidence against its existence as a nontrivial theory,
and its absence in our theory is therefore good news.  

Inverting (\ref{alphamu}), let us
express the bare coupling in terms of the effective coupling
 $\alpha = \alpha_0$ at zero momentum transfer $\mu = 0$ 
to this order as
\begin{align}
  \alpha_\infty = {\alpha \over \displaystyle{1 - {\alpha\over3\pi}
      \,\ln {M^2\over m^2}}}. \label{alphalambda}
\end{align} 
 To
investigate the plausibility of having a small bare coupling 
$\alpha_\Lambda$ in a model motivated by the QED sector of
our world, we take
$$
  \alpha \sim {1\over 137}
$$
and  $m \sim 0.511\,\textrm{MeV}$ to be of the order of the electron mass.  
Even if the mass $M$ of the ghost is a large as the mass-energy of the
visible part of our universe,
 for which we use, for the sake of the argument, 
the estimate
$$
M \sim 1.25\times 10^{82}\, \textrm{MeV},
$$
 we still find a bare coupling 
$$
  \alpha_\infty \sim {1\over 100},
$$
which is indeed small.  In addition, we find that
\begin{align}
  {\alpha_\infty \over 3\pi}\, \ln{M^2\over m^2} \sim 0.4161 < 1,
\label{firstorder}
\end{align}
which justifies our prior performing of the geometric sum to obtain 
(\ref{alphamu}).  The smallness of the first-order
$\alpha_\infty$ leaves plenty of room for possible
higher-order contributions.  For example,  
the second order contribution can be calculated and gives
 \begin{align}
  \alpha_\infty = {\alpha \over \displaystyle{1 - {\alpha\over3\pi}
      \,\ln {M^2\over m^2} - {\alpha^2\over4\pi^2}
      \,\ln {M^2\over m^2}}}.
\end{align}
so that 
$$
  {\alpha_\infty \over 3\pi}\, \ln{M^2\over m^2} \sim 0.4164,
$$
which is only a $7/10000$ correction to the first order result.

The evidence, to this order, is therefore that 
the the theory has an ultraviolet fixed point at $\alpha_\infty \sim 1/100$, 
and that the effective
coupling constant at any finite energy is smaller than this value.
We can therefore trust the accuracy of the perturbative calculation 
that we performed to produce this evidence.  This makes it 
feasible that the theory is indeed asymptotically safe.
Further support for this assertion will be given by an all-order
argument below.

Note that, since we could directly sum the geometric series, we
did not need to rely on a renormalization group equation to arrive
at the result for $\alpha_\mu$.  Our  result does reproduce the 
familiar first-order solution 
to the QED renormalization group equation  in the range 
$$
m^2 \ll \mu^2, {\mu'}^2 \ll M^2
$$
by using (\ref{alphamu}) to write $\alpha_\mu$ and $\alpha_{\mu'}$
 in terms of $\alpha_\infty$ and eliminating the latter to find the 
expected
\begin{align}
  \alpha_{\mu'} = {\alpha_\mu \over \displaystyle{1 - {\alpha_\mu\over3\pi}
      \,\ln {{\mu'}^2\over \mu^2} + o(\alpha_\mu^2)}}.
\end{align}
In this range, the $\beta$-function coincides with that of 
ordinary electrodynamics.  However, it is clear from 
(\ref{alphamu}) that the $\beta$-function gets modified when 
$\mu$ becomes comparable to the ghost mass $M$, where $\beta$ becomes 
explicitly dependent on $\mu$.  However, we can state that, to this order, 
$$
  \beta > 0
$$
on the whole range of positive $\mu$, and that
$$
  \beta \to 0 
$$
asymptotically as $\mu \to \infty$, as one would expect in an asymptotically 
safe theory.   

Let us consider whether higher-order corrections might spoil
the argument for 
asymptotic safety. A plausibility argument that the 
theory should remain asymptotically safe to all orders in perturbation 
theory can be based on the observation that all perturbative contributions
to the vacuum polarization are finite, just like
the first order (\ref{Pimunu0}).  This follows from 
 the known fact that the powers of 
$$
  \ln \Lambda
$$
appearing in the vacuum polarization $\Pi(q^2)$ come from 
internal Dirac loops only.   This is a consequence of gauge 
invariance.\footnote{For example, to second order there is a single 
Dirac loop, and indeed double logarithms
from self-energy and vertex sub-divergences cancel because, as a consequence
of gauge invariance,
 the corresponding couterterms occur with 
the same coefficient $(Z_\Lambda - 1)$ in the bare action (\ref{action1}).   
}  
Since each Dirac fermion loop can be individually replaced by a Dirac boson
ghost
loop of opposite sign, the $\Lambda$-dependent terms in a 1PI correction 
of arbitrary order
are proportional to
$$
  \left ( \sum_i c_i\,(\ln \Lambda)\right)^n = 0,
$$ 
so that the result is independent of the cutoff $\Lambda$.

Since $\Pi(q^2)$ is independent of $\Lambda$,
the bare parameter $\alpha_\Lambda$ still does not need
any 
infinite perturbative renormalization as $\Lambda \to \infty$ and can, as
before, be chosen as a constant 
independent of $\Lambda$.  We conclude that 
 perturbation theory, at least to the extent  that it remains 
accurate given that it is an asymptotic series, should not
spoil asymptotic safety.

\section{Mass and field renormalizations}
\label{sectionmass}

In the previous section we saw that corrections to the 
vacuum polarization  are finite due to cancellations between 
ordinary fermion and ghost loops.  On the other hand, self-energy 
diagrams are not finite without $\Lambda$-dependent mass and field 
renormalization counter-terms.   

Since the action is real-valued, the associated Hamiltonian, which acts on
an indefinite inner product space, is pseudo-hermitian.  
Pseudo-hermitian operators may have, in addition to to real 
eigenvalues, complex eigenvalues occurring in conjugate pairs and
associated to null eigenmodes \cite{bognar, malcev}.  In other words, 
as the coupling constant increases from $0$, a single field of real
mass may split into a pair of degrees of freedom 
of complex conjugate masses.  
It should be clear that the original field can only supply one
local counter-term to renormalize the common real part of these masses, so 
that their imaginary part, if any, is not an independent parameter.

\textit{We shall assume that real-valued mass counter-terms have been added 
so that the parameters $m = m_0$ and $M = m_1$ in the action are the 
real parts of the positions of the 
physical one-particle poles of the full propagator.  
This can always be done while
maintaining reality of the action, and therefore pseudo-hermiticity.}

Complex energies typically occur in cases where a positive-metric and a 
negative-metric state in the free theory are close compared to the scale  of 
their mixing term in the Hamiltonian.    
This can be understood in the elementary example of a two-state
Hamiltonian of the form 
$$
  H = \left(  
    \begin{array}{cc}
       \omega_1 & \gamma \\
      -\gamma & \omega_2
    \end{array}
    \right),
$$
which
 is pseudo-hermitian with respect to the
 inner product $\eta = \diag (1, -1)$, 
and whose eigenvalues become 
complex when 
$$
  {|\omega_1 - \omega_2|} < 2\gamma.
$$
In perturbation theory, this effect is a result of 
summation over virtual negative-metric intermediate 
states that displaces the energy poles into the complex plane
and requires a modification of the usual integration 
contours in the energy plane.  To avoid creating the 
impression that the contours are ad-hoc, let us 
derive the correct contour from first principles 
by applying perturbation theory to
 the above example, for simplicity taking 
$\omega_1 = \omega_2 \equiv \omega$.   We write
\begin{align*}
  \theta (t)\,\bra{1} e^{-iHt} \ket{1}
   &\equiv \int_\Gamma {dE\over 2\pi}\, e^{-iEt}\, f(E).
\end{align*}  
where $\Gamma$ indicates a deformation of the real line to be 
determined.  
As described in reference \cite{indef}, 
both $\Gamma$ and the function $f(E)$ may be computed by performing
perturbation theory in $t$-space to obtain the convergent series 
(here $V_{12} =  -V_{21}^* = \gamma$)
\begin{align}
&\int_{-\infty}^\infty {dE\over 2\pi}\, e^{-iEt}\,
  {i\over E-\omega + i0}\nonumber \\
&\quad   + (-i)^2  \int_{-\infty}^\infty {dE\over 2\pi}\,e^{-iEt}\,V_{12}V_{21}\, 
  \left({i\over E-\omega + i0}\right)^3  \nonumber\\
&\quad
    + (-i)^4 \int_{-\infty}^\infty {dE\over 2\pi}\, e^{-iEt}\,
  V_{12}V_{21}V_{12}V_{21} 
\left(i\over E-\omega + i0\right)^5
     + \cdots \nonumber \\
&= \theta(t)\, e^{-i\omega t} \left(
   1 + { V_{12}V_{21} (it)^2\over 2!} + {(V_{12}V_{21})^2\,(it)^4\over 4!} + \cdots\right)
  \nonumber\\
&= \theta(t)\, e^{-i\omega t} \cosh \gamma t \nonumber\\
&= \theta(t)\, \half\,\left(
e^{-i\,(\omega + i\gamma) t}+ e^{-i\,(\omega - i\gamma) t}
\right)
     \nonumber\\
  &= \int_\Gamma {dE\over 2\pi}\, e^{-iEt}\,
{i\over E - \omega - \displaystyle{V_{12}V_{21} / (E - \omega)}}, \label{tpert}
\end{align}
where $\Gamma$ is obtained by a deformation of the real line into the 
complex plane to
run above both poles of 
$$ 
f(E) \equiv {i\over E - \omega + \displaystyle{\gamma^2 / (E - \omega)}}
$$ 
including when the poles are away from the real axis in the complex
plane.  It is important to note that we have to \textit{first} 
perform the $E$-integrals in the first line 
and then sum the series \cite{indef}.    
 Note that 
the order of integration and summation 
cannot be 
interchanged, since the momentum-space geometric series under the 
integral does not converge for all $s$.  
Thus, the perturbative series can be summed   in position 
space where it converges, but 
we cannot expect to get the correct answers by formally
summing the divergent Fourier transformed terms  in momentum space.  
For further examples where 
such integration contours
are unambiguously determined in a non-perturbative framework,
see reference \cite{indef}.

Notice, though, that the functional form of $f(E)$, though not 
the contour $\Gamma$, 
could have been obtained directly from a formal geometric sum
in momentum space.\footnote
{
It is important to note, however, that the propagator is a distribution
$\int_\Gamma dE\, f(E)\,(\,\cdots)$, which is 
specified not only by the function $f(E)$ but also by the 
contour $\Gamma$.  
Momentum-space perturbation theory formally gives the function $f(E)$, but 
not the contour, and thus does not fully specify the distribution.  
Knowing the contour becomes essential when
we use $f(E)$ as an internal line in a higher order diagram, so that an
integration over $E$ is required.  See reference \cite{leewick} for an 
example where using the wrong distribution for an internal line
gives incorrect results.  Reference \cite{indef} contains a discussion 
on how to calculate such diagrams unambiguously in a non-perturbatively well-defined
approach.  
}

For comparison with field theory, notice that
the poles of $f(E)$, which are the energy eigenvalues of $H$, acquire imaginary
 parts as a consequence
of the fact that the imaginary part of the contribution to the denominator 
of $f(E)$,
$$
  \textrm{Im}\, {-V_{12}V_{21}\over E- \omega} 
$$
is negative above the real axis due to
$V_{12} = -V_{21}^*$, which is the case for matrix elements between states 
of opposite metric.  As a result, this term 
will cancel the imaginary part of the other contribution $E-\omega$
to the denominator of $f(E)$ somewhere in the upper half plane, and we
get a complex pole as long as the real parts of these two terms
also cancel which,
in the non-degenerate case where $\omega_1 \ne \omega_2$,
will happen if   $\omega_1$ is close  to $\omega_2$
compared to the scale set by $\gamma$.

This generalizes to field theory as follows.  Consider doing perturbation
theory based on a  Dirac particle of rest mass $m$. 
The propagator is
$$
  {i\over \slashiv p - m - \Sigma (\slashiv p)},
$$
where $-i\Sigma$ denotes a sum of diagrams that are irreducible with respect 
to the particle whose self-energy we are determining.  
Again, a full specification of the distribution consists not only 
of this function but also the associated integration contour, which can be
obtained through a careful calculation in the spirit of (\ref{tpert}). 
The irreducible contribution to the 
forward scattering amplitude for 
the Dirac particle is 
$$
  \mathcal M (p, s \to p, s) = - Z\, \bar u^s_p\,\Sigma\,u^s_p.
$$
This is independent of the spin $s$ by rotational invariance, and is the complex
conjugate of the corresponding anti-particle matrix element.   
With respect to the conjugation defined by
$\bar \Sigma \equiv \gamma^0 \,\Sigma^\dagger \,\gamma^0$, 
the optical theorem may be written as
\begin{align*}
 \textrm{Im}\, (-\Sigma) &= {1\over 2mZ} \,\textrm {Im}\,\mathcal M (p, s \to p, s) \cdot \mathbf{1}
   \\
   &= {1\over 2mZ} \left(\half\right) \int d\mu_f\, 
         \mathcal M (p, s \to f)\, \mathcal M (f \to p,s)
      \,\delta^4 (p_f - p),
\end{align*}
where $d\mu_f$ is the appropriate Lorentz-invariant measure on the 
space of all possible final states that may be obtained by cutting 
irreducible diagrams.

When $\Sigma(\slashiv p)$ has a multi-particle cut starting 
at $p^2 < m^2$, the particle is unstable and can  decay to the
multi-particle states.  Consider the propagator
$$
  {i\over \slashiv p - m - \Sigma (\slashiv p)}.
$$ 
We see that if $\textrm{Im}\, (-\Sigma)$ has a cut starting at some
$p^2 < m^2$, and is positive above and negative below the cut, the propagator 
will have no pole on the physical sheet.  It will have a pole with 
nonzero negative  imaginary part on the unphysical second sheet.
  We no longer have a particle but instead
a resonance.  

In ordinary hermitian quantum field theories, 
$\textrm{Im}\, (-\Sigma)$ will always be positive, and the resonance 
poles will not be be on the physical sheet.  In pseudo-hermitian
theories, however, we have
$$
\mathcal M (p, s \to f)\, \mathcal M (f \to p,s)
= - |\mathcal M (p, s \to f)|^2 < 0,
$$
whenever the initial and final states are non-null states of
 opposite metric.  As a result,
$\textrm{Im}\, (-\Sigma)$ will become negative if this contribution 
is larger than the total contribution of decays to states of the same 
metric.  We find that in this case, the 
poles of 
$$
  {i\over \slashiv p - m - \Sigma (\slashiv p)},
$$
will be on the \textit{physical} sheet, at complex conjugate 
positions with nonzero imaginary parts \cite{leewick}.  
Such poles are not 
 resonances.  Instead, they 
correspond to a pair of null energy eigenstates 
 with complex conjugate rest masses.  Such conjugate
null states are dual, i.e., the inner product on the subspace spanned
by these states can be
brought to the form 
$$
\left(\begin{array}{cc}0&1\\1&0\end{array}\right).
$$
\textit{In other words, whenever a particle is unstable with respect to 
decay to a multi-particle state of opposite metric, it may split into 
a pair of degrees of freedom of complex conjugate masses.}  
 
Let us consider whether the masses of the two types of 
Dirac particles in
our theory may become complex.  
\textit{We shall assume, for the rest of this paper, that 
both $m$ and $M$ are nonzero, and that $M \gg m$.}

 First we consider the ghost.  
The lowest order negative 
contribution to $\textrm{Im}\, (-\Sigma)$
in the ghost forward amplitude is given by the indicated vertical 
cut in the diagram 
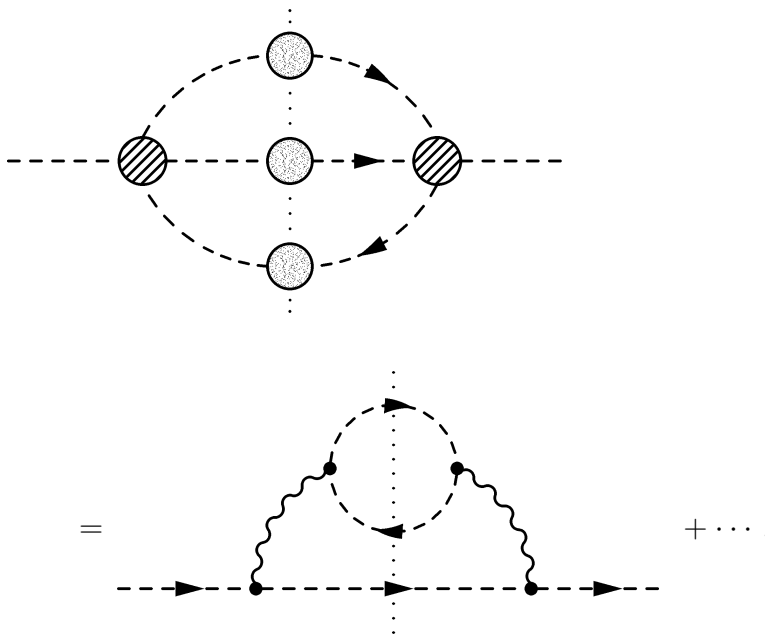
\newenvironment{pp0}
{\begin{fmfgraph*}(75,45)
    \fmfleft{a,p,inv,c}\fmfright{aa,pp,invv,cc}
  }
  {\end{fmfgraph*}}
\begin{align*}
&\parbox{75mm}
{
\begin{fmfgraph}(75,40)
\fmfleft{l}
\fmfright{r} 
\fmftop{t}
\fmfbottom{b}
\fmf{dots,background}{t,v2,v3,v4,b}
\fmfrpolyn{shaded,smooth}{G}{4} 
\fmfpolyn{shaded,smooth}{K}{4}
 \fmf{dashes}{l,G1}
 \fmf{dashes}{K1,r} 
\fmf{dashes,tension=.3}{G3,v3}
\fmf{scalar,tension=.3}{v3,K3}
\fmf{dashes,left=.3,tension=.25}{G2,v2} 
\fmf{scalar,left=.3,tension=.25}{v2,K2} 
\fmf{scalar,left=.3,tension=.25}{K4,v4} 
\fmf{dashes,left=.3,tension=.25}{v4,G4} 
\fmfv{decor.shape=circle,decor.filled=10,decor.size=.08w}{v2,v3,v4}
\fmffixed{(0,.7h)}{v4,v2}
\end{fmfgraph}
} \\
&\qquad=
\parbox{75mm}
{
{\begin{pp0}
  \fmf{phantom,tension=.2}{a,av,aa}
  \fmf{phantom}{c,cv,cc}
  \fmf{dots,tension=.1}{av,cv}
  \fmf{scalar}{p,v1}  
  \fmf{scalar,tension=.5}{v1,v2}
  \fmf{scalar}{v2,pp}
  \fmf{phantom,tension=.3}{inv,v3}
  \fmf{phantom,tension=.5}{v3,v4}
  \fmf{phantom,tension=.3}{v4,invv}
  \fmffreeze
  \fmf{boson,left=.3}{v1,v3}
  \fmf{scalar,left,tension=.6}{v3,v4,v3}
  \fmf{boson,left=.3}{v4,v2}
  \fmfdot{v1,v2,v3,v4}
\end{pp0}} 
}+ \cdots,
\end{align*}
since states containing even an odd numbers of ghost  anti-particles 
have  opposite sign  metric.  

This cut starts at $p^2 = (3M)^2$, which is well-separated from $M^2$.  
Assuming $\textrm{Re}\, \Sigma (\slashiv p = M) = 0$, which can always be 
ensured by including an appropriate real counter-term in the bare action, 
the mass pole remains at the real 
position $p^2 = M^2$.  Indeed, the indicated decay is prevented by 
kinematics, and so does not lead to an instability and an associated complex
mass.  

The same goes for the lowest order negative 
contribution to $\textrm{Im}\, (-\Sigma)$ in  the fermion forward 
scattering amplitude
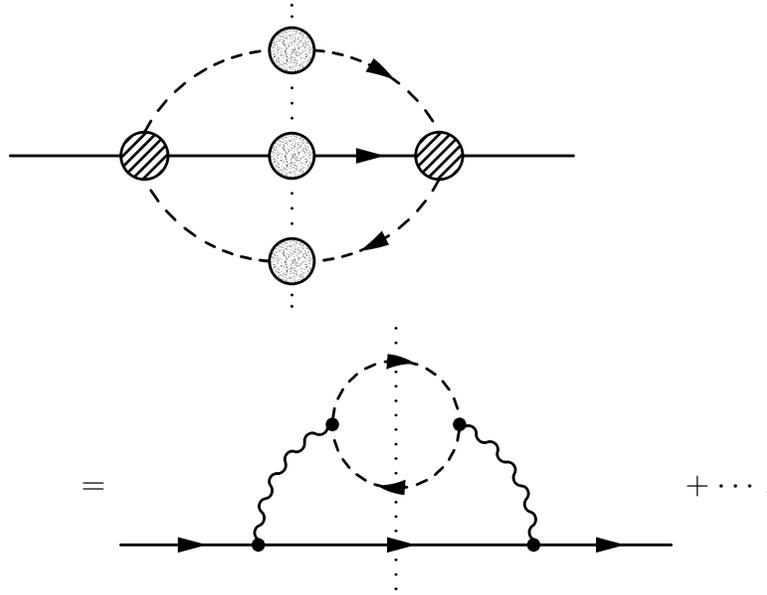
\begin{align*}
&\parbox{75mm}
{
\begin{fmfgraph*}(75,40)
\fmfleft{l}
\fmfright{r} 
\fmftop{t}
\fmfbottom{b}
\fmf{dots,background}{t,v2,v3,v4,b}
\fmfrpolyn{shaded,smooth}{G}{4} 
\fmfpolyn{shaded,smooth}{K}{4}
 \fmf{plain}{l,G1}
 \fmf{plain}{K1,r} 
\fmf{plain,tension=.3}{G3,v3}
\fmf{fermion,tension=.3}{v3,K3}
\fmf{dashes,left=.3,tension=.25}{G2,v2} 
\fmf{scalar,left=.3,tension=.25}{v2,K2} 
\fmf{scalar,left=.3,tension=.25}{K4,v4} 
\fmf{dashes,left=.3,tension=.25}{v4,G4} 
\fmfv{decor.shape=circle,decor.filled=10,decor.size=.08w}{v2,v3,v4}
\fmffixed{(0,.7h)}{v4,v2}
\end{fmfgraph*}
} \\
&\qquad=
\parbox{75mm}
{
{\begin{pp0}
  \fmf{phantom,tension=.2}{a,av,aa}
  \fmf{phantom}{c,cv,cc}
  \fmf{dots,tension=.1}{av,cv}
  \fmf{fermion}{p,v1}  
  \fmf{fermion,tension=.5}{v1,v2}
  \fmf{fermion}{v2,pp}
  \fmf{phantom,tension=.3}{inv,v3}
  \fmf{phantom,tension=.5}{v3,v4}
  \fmf{phantom,tension=.3}{v4,invv}
  \fmffreeze
  \fmf{boson,left=.3}{v1,v3}
  \fmf{scalar,left,tension=.6}{v3,v4,v3}
  \fmf{boson,left=.3}{v4,v2}
  \fmfdot{v1,v2,v3,v4}
\end{pp0}}
}
+ \cdots,
\end{align*}
where the indicated decay is, once again, prevented by kinematics
due the the assumption  $M \gg m$. 
The fermion, like the Dirac boson, is stable and its mass remains real.  

We conclude that the masses of the fundamental Dirac fermion and 
Dirac bosonic ghost 
remain real under perturbative corrections.  

There are, however, composite degrees of freedom 
whose masses we do expect to become complex.  The lightest of 
these is the
ghost  analogue of the singlet ground state of
positronium.  The singlet state of ordinary 
positronium  is an unstable state consisting of a fermionic
particle and anti-particle in the singlet configuration. 
This state is not part of the spectrum but corresponds to
a resonance.  
The ghost analogue 
consists of a ghost particle and ghost anti-particle in
the singlet configuration.  In this case, however, the
instability of the state under decay to states of 
opposite metric will split it into a 
pair of null degrees of freedom
of complex conjugate masses.  The corresponding
poles will be on the physical sheet, and 
will appear, for example, in the 
full vacuum polarization 
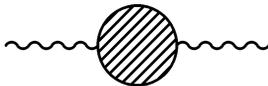
\begin{align*}
\begin{fmfgraph*}(35,25)
\fmfleft{a}
\fmfright{b}
\fmf{photon}{a,v,b}
\fmfblob{.3w}{v}
\end{fmfgraph*}
\end{align*}
Since the singlet is a scalar, we expect a contribution
 to the vacuum polarization of the approximate form
$$
   {i C\over p^2 - \mu^2 - \Sigma(p)},
$$
where $C$ is a constant and 
where $\Sigma$ is again the irreducible forward scattering amplitude
for the effective ghost positronium degree of freedom.  
As in the Dirac case, cuts corresponding to intermediate states of 
opposite metric will
give negative contributions to $\textrm{Im}\, (-\Sigma)$ , which
may then similarly cause the pole to split and move to complex
conjugate positions away from the real axis on the physical
 sheet.\footnote
{Notice that singlet ghost positronium is 
a \textit{bosonic} Klein-Gordon degree of freedom  with 
\textit{negative}
 metric.
If such a field had appeared in the original Lagrangian, it would
have violated micro-causality \cite{bogolubov}. 
For describing a
composite object, it is acceptable.
}

The
parameter $\mu$ above is approximately 
 the real part of the rest mass of the bound state,.
As in the case of ordinary positronium, this
 mass can be
approximated
by subtracting the non-relativistic binding energy to give,
for ghost positronium, the value
\begin{align}
  \mu \equiv 2M - \alpha^2M/4.    \label{realpos}
\end{align}
Then 
 $\textrm{Im}\, (-\Sigma)$
is determined by the various decay processes that may affect 
the composite degree of freedom.   
A first approximation may be obtained by 
considering the constituent particles at relative rest,
which translates to calculating certain simple
Feynman diagrams near the two-particle  threshold $p^2 = (2M)^2$ as 
described, for example, in
\cite{IZ}.  

For ghost positronium, the first relevant contribution 
to $\textrm{Im}\, (-\Sigma)$  is 
given by the vertical cut in the diagram 
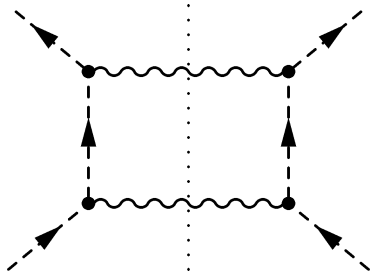
\begin{align*}
\begin{fmfchar*}(60,35)
  \fmfleft{em,ep}
  \fmfright{fb,f} 
  \fmf{phantom}{em,av,fb}
  \fmf{phantom}{ep,bv,f}
  \fmffreeze
  \fmf{dots}{av,bv}
  \fmf{scalar}{em,Zee}
  \fmf{scalar,tension=.5}{Zee,Zee1}
  \fmf{scalar}{Zee1,ep}
  \fmf{photon,tension=.4}{Zee,Zff}
  \fmf{photon,tension=.4}{Zee1,Zff1}
  \fmf{scalar}{fb,Zff}
  \fmf{scalar,tension=.5}{Zff,Zff1}
  \fmf{scalar}{Zff1,f}
  \fmfdot{Zee,Zff,Zee1,Zff1}
\end{fmfchar*}
\end{align*}
calculated at the threshold $p^2 = (2M)^2$ .
This cut corresponds to the decay of the composite state
to a pair of photons.  
The indicated 
cut starts at $p^2 = 0$ and contributes negatively to $\textrm{Im} (-\Sigma)$
along the real axis.
If negative contributions are dominant at the energy $\mu$ 
 (\ref{realpos}) , this contribution will make the
pole at $\mu^2$ split into a pair of complex conjugate poles away from
the real axis.  
To the same order, the following diagrams also contribute to 
$\textrm{Im} (-\Sigma)$.    
\begin{equation}
\parbox{60mm}
{
\begin{fmfchar*}(60,25)
  \fmfleft{em,ep} 
  \fmfright{fb,f} 
  \fmf{phantom}{em,av,fb}
  \fmf{phantom}{ep,bv,f}
  \fmffreeze
  \fmf{dots}{av,bv}
  \fmf{scalar}{em,Zee,ep}
  \fmf{photon}{Zee,v1}
  \fmf{photon}{v2,Zff}
  \fmf{fermion,left,tension=0.5}{v1,v2,v1} 
  \fmf{scalar}{fb,Zff,f}
  \fmfdot{Zee,Zff,v1,v2}
\end{fmfchar*}
}
+
\parbox{60mm}
{
\begin{fmfchar*}(60,25)
  \fmfleft{em,ep}
  \fmfright{fb,f}
  \fmf{phantom}{em,av,fb}
  \fmf{phantom}{ep,bv,f}
  \fmffreeze
  \fmf{dots}{av,bv}
  \fmf{scalar}{em,Zee,ep}
  \fmf{photon}{Zee,v1}
  \fmf{photon}{v2,Zff}
  \fmf{scalar,left,tension=0.5}{v1,v2,v1} 
  \fmf{scalar}{fb,Zff,f}
  \fmfdot{Zee,Zff,v1,v2}
\end{fmfchar*}
}\end{equation}
Due to the assumption, $m \ll M$, the threshold $p^2 = (2m)^2$  for creating two 
fermions is lower than $\mu^2$, 
so the
indicated cut in the first of these diagrams also corresponds to an
instability that contributes 
negatively to $\textrm{Im} (-\Sigma)$ in the vicinity of $\mu^2$.  
The cut in the 
second diagram starts to contribute positively to $\textrm{Im} (-\Sigma)$,
but only above the threshold $p^2 = (2M)^2$.  
Since $\mu < 2M$, the decay is kinematically ruled
out and does not affect our conclusion that we obtain complex conjugate
physical poles.   

In addition to these unstable states, we also expect stable bound states 
consisting of ghost-lepton atoms.   One consists of an electron and an 
anti-ghost and has negative metric.  Its anti-atom has 
positive metric.  These have approximate rest mass
$$
  M + m - {\alpha^2\over 2} \left({mM\over M + m}\right).
$$ 

The analytic structure of diagrams involving multi-particle 
intermediate states is much more problematic.  
For example, it has been
argued that diagrams involving
intermediate states of complex conjugate rest masses,
calculated in a Hamiltonian approach, have
non-covariant singularities that break Lorentz invariance
\cite{nakanishi, gleeson}.   
We will discuss this issue in 
section \ref{sectionlorentz}, where we propose a solution
based on carefully constructed Lorentz invariant multi-particle 
state spaces that are different from the spaces used by prior  authors.  
The resulting theory will be Lorentz-invariant by construction, 
and so will the analytic structure of Feynman diagrams. 

As an example, in our approach
we would expect a cut starting on the real axis at 
$$
  p^2 = \left\{2\left(2M - \alpha^2M/4\right)\right\}^2,
$$
corresponding to pairs of conjugate ghost positronium states.  However,
as we will discuss in the section on Lorentz invariance, 
when considered properly in a specific non-perturbatively well-defined
framework, interactions are expected to cause 
this cut to split into a pair of conjugate cuts ``infinitesimally''
above and below the complex plane.

Note that we will also have cuts
starting at complex branch points.  For example, two of these cuts,
corresponding to two-particle states consisting of an  electron and 
a ghost positronium, will start approximately at 
$$
  \left\{m + 2M - \alpha^2M/4 \pm i\gamma/2\right\}^2,  
$$
where $2M + \alpha^2M/4 \pm i\gamma/2$ are the complex
ghost positronium masses.  
For each cut, its complex conjugate mirror image also appears
as a consequence of the  pseudo-hermiticity of the Hamiltonian.

Since mass and field renormalizations are $\Lambda$-dependent, 
unlike the charge renormalization,
it is perhaps not obvious that they do not suffer from
Landau-pole type obstructions similar to those 
that appear in ordinary quantum 
electrodynamics.  We verify that such an obstruction is 
absent for the field renormalization
constant
$Z_{i\Lambda}$.  To second order, we have 
$$
  Z_{i\Lambda} = {1\over 1 + 
    \displaystyle{{\alpha_\Lambda\over 4\pi}
   \left (\ln {\Lambda^2 \over m_i^2} - 3 \ln 3 + {9 \over 4}\right)}}
$$  
in a gauge where infrared divergences vanish \cite{IZ}.  Unlike 
the case of the charge, this has an explicit dependence on the 
regularization $\Lambda$. 
In ordinary electrodynamics, we are prevented from taking the limit
$\Lambda \to \infty$, at least in perturbation theory,
due to the fact that $\alpha_\Lambda \to \infty$
for some finite value of $\Lambda$.  
In the present case, however, we have arranged for $\alpha_\Lambda$
to remain small as $\Lambda \to \infty$, so $Z_\Lambda$ 
remains nonzero for all $\Lambda$.  As a result, there is no
obstruction to removing the regularization.

\section{Unitarity and scattering}
\label{scattering}

The time evolution described by our action is pseudo-unitary.  
This means that inner products and normalizations are preserved 
by time evolution, as is the trace of the density matrix, which can 
be normalized to unity.  However, the density matrix is not positive definite, 
and does not have a conventional probability interpretation.  
 
However, Lee and Wick
noted that if all negative-definite  states are  ``unstable''
in the 
sense of acquiring non-real complex energies
due to interactions, we will obtain a 
unitary theory by constraining the 
state space to real-energy states \cite{leewick}.  This proposal 
worked well to eliminate single-ghost states, but was never 
satisfactorily generalized to the elimination of multi-ghost states.
Attempts by  previous authors to do so were either 
ad hoc, ambiguous,  and probably nonperturbatively ill-defined
 \cite{lee1, cutkosky, BG, nakanishiprime}, or broke Lorentz invariance
\cite{nakanishi, gleeson}.

We will argue that  the original real-energy  proposal
 can, with some care, be generalized 
to the elimination of multi-ghost states in a Lorentz invariant and 
non-perturbatively well-defined manner, and
is therefore
sufficient for  obtaining a unitary theory in the case of
ghost QED.  
 
\textit{
Specifically, we will argue in the following sections that
multi-ghost states can be quantized in a Lorentz invariant way.
At this point, their energies may be real, but 
taking into account interactions, the energies generically 
become
non-real as long  we impose a
 long-distance
cutoff.  We then apply the Lee-Wick real-energy 
constraint to eliminate these states first and only 
then remove the cutoff.
The resulting theory should be unitary.  
}

Therefore, in this section and the next, we review how the 
elimination of complex-energy states affects scattering 
calculations.  We essentially follow Lee and Wick \cite {leewick}.

Ordinary scattering theory assumes a free 
 Hamiltonian $H_0$ and an interacting hermitian Hamiltonian $H$, both
hermitian and with 
coinciding continuous spectra, and constructs operators $W_-$
and $W_+$ that allow us to obtain the generalized eigenstates 
of the interacting Hamiltonian from those of the free Hamiltonian as 
$$
  \ket{E_\alpha,\alpha,\textrm{in}}_{H} = W_- \ket{E_\alpha,\alpha}_{H_0}, 
  \qquad
  {}_H\bra{E_\alpha,\alpha,\textrm{out}} = {}_{H_0}\bra{E_\alpha,\alpha}W_+^{\dagger},
$$
where $\alpha$ stands for any additional quantum numbers.
Given
these two operators, the scattering matrix can be written as
$$
  S = W_+^\dagger W_-.
$$
To obtain a probability interpretation in an indefinite theory, we need to 
restrict the state space to a physical subspace that is positive definite
and invariant under time evolution.  
Since any complex-energy eigenstates have zero norm, we may follow Lee and
Wick and first eliminate these
by constraining the physical subspace to lie within the real
eigenvalue spectrum of $H$.  

Let us discuss
the implications of this constraint for scattering.  We assume that the free Hamiltonian $H_0$ has real spectrum and we assume, as before,
that the parameters in $H_0$ have been chosen so that, for 
a subset of quantum numbers, its spectrum coincides with the real part of the
interacting spectrum.  The Lippmann-Schwinger equations may be used to formally
represent $W_{\pm}$ as 
$$
  W_{-} \ket{E,\cdots}_{H_0}  =  \left(1 + {1\over E - H + i\epsilon}\,(H-H_0) \right)\ket{E,\cdots}_{H_0},
$$
with the hermitian conjugate equation for $W_+^\dagger$.  

In an indefinite theory, the Lippmann-Schwinger equations 
remain sensible for the states of real $E$, but 
the interpretation of the S-matrix in terms of a scattering process runs into 
the following hurdle:  Ordinarily, the operators $W_\pm$ may be also
obtained as weak limits
\begin{align*}
  W_- &= \lim_{t\to\infty} e^{-iHt}e^{iH_0 t}, \\
  W_+^\dagger &= \lim_{t\to\infty} e^{iH_0t}e^{-iH t},
\end{align*}
from which the physical scattering interpretation may be derived.  
However, if $H$
has complex eigenvalues with positive imaginary part, these limits 
will not in general exist. 
Simply stated, not all free eigenstates
of real energy are orthogonal to the complex energy eigenstates of the 
interacting $H$, so that applying $e^{-iHt}$ to a free eigenstate will 
give an exponentially diverging contribution as $t \to \infty$.  

The solution proposed by Lee and 
Wick is to make sure that we project the complex energy null states,
which are not in the physical subspace, out of the 
state  $e^{iH_0 t} \ket{E_\alpha,\alpha}_{H_0}$ at time $-t$, and 
instead use as initial state
$$
  e^{iH_0 t} \ket{E_\alpha,\alpha}_{H_0} - \int_{E_\beta\notin\mathbf{R}
} d\beta\,\ket{E_\beta,\beta}_H\bra{E^*_\beta,\beta}
   e^{iH_0 t}\ket{E_\alpha}_{H_0}. 
$$
The projection depends on the solution to the full dynamics of  
$H$ and can be calculated order by order in perturbation theory.

Equivalently, we have to apply $e^{-iHt}e^{iH_0 t}$ not 
to the free eigenstate $\ket{E_\alpha,\alpha}_{H_0}$ but instead 
to the state
\begin{align}
& \ket{E_\alpha,\alpha}_{H_0}^\text{p}\nonumber \\
&\quad=\ket{E_\alpha,\alpha}_{H_0} - \int_{E_\beta
\notin\mathbf{R}} d\beta\,e^{iE_\alpha t}e^{-iH_0t}\ket{E_\beta,\beta}_H\braket{E^*_\beta,\beta}{E_\alpha}_{H_0}\nonumber \\
&\quad =
  \ket{E_\alpha,\alpha}_{H_0} \nonumber\\ 
&\qquad \quad- 
\int_{E_\beta\notin\mathbf{R}} d\beta\int_\mathbf{R} d\gamma\,e^{i(E_\alpha - E_\gamma) t}\ket{E_\gamma,\gamma}_{H_0}\braket{E_\gamma, \gamma}{E_\beta,\beta}_H\braket{E^*_\beta,\beta}{E_\alpha}_{H_0}.
\label{correlation}
\end{align}
Interestingly, we can argue that the second term vanishes in the 
limit $t\to\infty$ that interests us, but only as far as a local observer
in the far past is concerned.  A careful way of doing this 
is by considering a smooth wave packet
\begin{align*}
& \int d\alpha \,f(\alpha)\ket{E_\alpha,\alpha}_{H_0}^p \\
&\quad =
 \int d\alpha \, f(\alpha)\ket{E_\alpha,\alpha}_{H_0}\\
&\qquad - \int_{E_\beta\notin\mathbf{R}} d\beta\int_\mathbf{R} 
d\gamma \ket{E_\gamma,\gamma}_{H_0}
\braket{E_\gamma, \gamma}{E_\beta,\beta}_H \times \nonumber \\
&\qquad\qquad\qquad \qquad\times\int d\alpha \,f(\alpha)
\,e^{i(E_\alpha - E_\gamma) t}\braket{E^*_\beta,\beta}{E_\alpha, \alpha}_{H_0}.
\end{align*}
Assuming ${}_H\braket{E^*_\beta,\beta}{E_\alpha, \alpha}_{H_0}$ is sufficiently smooth -- which is expected to be the case since the real $E_\alpha$ never concides with the complex $E_\beta$ -- we
use the Riemann-Lebesgue lemma to conclude that the last integral goes to 
zero polynomially as $e^{-iE_\gamma t}/t^a$
when $t\to\infty$.  In other words, 
$$
   \int d\alpha \, f(\alpha)\ket{E_\alpha,\alpha}_{H_0}^p \to 
    \int d\alpha \, f(\alpha)\ket{E_\alpha,\alpha}_{H_0}
$$
as $t\to\infty$.  
This equation expresses the interesting property  that
two states, known to have very different time evolution, are locally
indistinguishable to arbitrary precision as $t\to\infty$ in the 
usual topology \cite{bognar}.  
Specifically, the state on the left remains bounded when we apply
$e^{-iHt}e^{iH_0 t}$ to it, whereas the 
one on the right depends exponentially on $t$
due to the factor
$e^{-iHt}$.   The reason for this
is that the local distinguishability of the states decreases 
polynomially in the far past, and a polynomial decrease will always 
be overcome by the exponential 
growth in the time evolution.

This means that an experimenter far enough in the past  cannot locally
distinguish the input states
that he or she is preparing from free eigenstates of $H_0$. 
Although the prepared states unavoidably contain 
nonlocal 
correlations encoded  in the second term in equation (\ref{correlation}),
these correlations are not locally detectable to an experimenter located far
enough in the past that
the wavelength of these correlations is smaller than the experimental resolution.
In contrast to ordinary scattering theory where 
polynomially decaying terms can be discarded in the limit,  here
the  correlations represented by the extra terms are not innocuous
and have to be kept, since
they 
are exponentially magnified by the time evolution as the input states evolve towards 
the scattering region, and their presence cancels out the exponential 
divergence of the state that would otherwise have occurred.  

Since the correlations in the initial state are not locally detectable 
in the sufficiently far past,
their effect unfolding as time goes on may give the appearance of 
acausality, as has been discussed by previous authors. 
 It should be clear from the formalism that there is 
no real acausality, since the information, called ``precursors'' by Lee,
is in fact present in the initial state (\ref{correlation}).  
We discuss the issue of causality more fully in  section
\ref{sectioncausality}.

\section{Calculation of the S-matrix}

Given the rather complex characterization (\ref{correlation}) of 
the asymptotic states, how does 
one in fact calculate physical scattering amplitudes 
perturbatively in a way that
properly takes into account the rather complicated precursor information
so as to avoid exponential blow-up of the S-matrix?  

As explained  by Lee and Wick \cite{leewick},
 it is straightforward in principle to  calculate amplitudes
in the reduced theory obtained by dropping the non-real energy 
 states from the spectrum.
One proceeds by calculating amplitudes without this restriction, as one 
would in ordinary quantum field theory, 
between states $\ket{p_\alpha,\alpha}_{H_0}$.  
One does this on a large but finite time interval $2t$.  As we
discussed, the resulting S-matrix elements contain terms that 
diverge exponentially as $t\to\infty$.   After identifying and discarding 
these terms, Lee and Wick showed that we obtain exactly the desired
S-matrix elements between the states $\ket{p_\alpha,\alpha}^\text{p}_{H_0}$,
expressed in equation (\ref{correlation}), 
of the constrained  theory without the non-real energy  states.

In practice, the exponentially growing terms may often
be discarded via the following trick.  If the state depends 
analytically on the various 
scattering invariants,
scattering amplitudes in the full unconstrained theory
can often be expressed as contour integrals over the  space of 
complexified invariants.  Discarding exponentially 
growing terms from scattering amplitudes 
can then often  be implemented
in a straightforward way by modifying the
contour integrals on the complex plane.  
For example, for a simple two-particle scattering, this  can be done by 
taking the integration contour in the $s$-plane to be along the
real line instead of the contour that runs  above the non-analyticities
in the upper complex plane representing exponentially growing 
contributions of  complex energy states in the full theory.   

Useful as such contour tricks are, they are not foundational
to the definition of the theory.  The Lee-Wick real-energy 
constraint is non-perturbatively well-defined  and 
does not assume any particular analytic properties  of 
amplitudes.

\section{Lorentz invariance}
\label{sectionlorentz}

In its Lagrangian formulation, the theory appears manifestly Lorentz-invariant.  
We shall argue that the theory can be quantized in a way that preserves this Lorentz
invariance even after applying the Lee-Wick real energy constraint.  Our construction 
 will be
in principle non-perturbatively well-defined.

This assertion may seem to be in conflict
 with prior work by Nakanishi \cite{nakanishi} and Gleeson et al.\
\cite{gleeson}.  These authors argued that, if the unconstrained theory has 
complex mass
states, certain amplitudes where such states appear in loops lose
Lorentz invariance after 
applying the Lee-Wick real energy constraint.  
They  based their calculations on the Hamiltonian 
approach 
originally advocated by Lee \cite{lee2}, 
which is in principle non-perturbatively well-defined.
The fact
 that they ran into problems should therefore be taken very seriously.\footnote{
An alternative approach to obtaining a unitary 
theory 
was investigated by Cutkosky et al \cite{cutkosky}.  These authors proposed modifying
diagrams order by order according to heuristic prescriptions based on unitarity 
considerations.  The resulting amplitudes are Lorentz-invariant, but as was
 noted by these 
authors themselves, their prescription was at best incomplete due to unresolved
ambiguities.  As was argued also, amongst others,
 by Nakanishi \cite{nakanishi} and
Boulware and Gross \cite{BG}, the prescription may not have 
non-perturbative meaning.  For example, it would appear that no Hamiltonian can 
reproduce their prescription.

For these reasons, we shall not discuss  the approach of Cutkosky et al. further.
Instead, we shall
insist on sticking to calculations that are
in principle non-perturbatively well-defined, based either on path integral or 
Hamiltonian arguments.}

We will argue that the problem pointed out by Nakanishi and 
by Gleeson et al.\ 
can be overcome by basing the state space on a class of
distributions studied by Gel'fand and Shilov \cite{gelfand1, gelfand2} . 
Our construction will allow us to define the multi-ghost state spaces
and the Lee-Wick real energy constraint 
in a Lorentz-invariant way.   Our method is in principle
non-perturbatively well-defined.  

Let us discuss  the problem in more detail  
by way of an example.
Consider a theory that contains ghost  scalar  particles  of complex
conjugate masses $M$ and $M^*$ (for example, the two ghost positronium states
considered previously).  Next consider a one-loop diagram whose
intermediate two-particle state consists of an $M$-particle and an $M^*$-particle.
Writing their energies as $\sqrt{M^2 + \bar k^2}$ and $\sqrt{{M^*}^2 + 
({\bar p-\bar k})^2}$ and assuming, as these previous authors did, 
 real space-like momenta $\bar k$ and $\bar p$, the description 
of the two-ghost state space is \textit{not} Lorentz invariant, and 
the resulting diagram contains a fish-shaped non-analytic 
region in the $s = p^2$ plane \cite{nakanishi, gleeson},
whose
 shape depends  on the Lorentz frame.\footnote{
The non-invariance of the two-ghost state space will 
nevertheless not be visible
 in the unconstrained theory 
as long as the external legs are non-ghost particles of real mass
with asymptotic wave packets  that are 
analytic at least on this fish-shaped non-analyticity region in  $s$.
 In this case,
scattering amplitudes in the unconstrained theory between non-ghost 
states will be
Lorentz invariant.
This is because the integration path for calculating scattering amplitudes 
in the unconstrained theory does not 
run along the real line in the $s= p^2$ plane, 
but instead runs above the region of non-analyticity, as
we saw in a toy example in section \ref{sectionmass}.
So if the  asymptotic wave packets  are 
chosen, for example,  to be holomorphic in $s$,
we can move the contour down towards the real line while continuing the 
integrand into the original non-analytic region.  
This analytically continued integrand turns out to be
 a \textit{Lorentz invariant} function
with a cut along the real axis starting at 
$(M+M^*)^2$ \cite{lee2, gleeson}.  Thus,
provided the asymptotic wave packets are holomorphic,
 the scattering amplitude in question is in fact Lorentz invariant.}

 The apparent conflict  with Lorentz invariance
occurs when we apply the Lee-Wick real energy constraint 
to  obtain a unitary theory.  
Given the real-momentum characterization of the 
two-ghost states described in
the previous paragraph, it is clear that, apart from a set of measure zero, the 
two-ghost intermediate states have complex energies.
As we discussed, Lee and Wick showed that removing  complex-energy states
translates  into discarding exponentially 
growing terms from the scattering amplitudes \cite{leewick}, and  that 
this can be done in simple cases such as this one by 
taking the integration contour in the $s$-plane to be along the
real line instead of running above the non-analyticities.
Applying the 
Lee-Wick trick, the modified contour along the real line will 
bisect the non-invariant region of  non-analyticity, and the
scattering amplitude will not be Lorentz invariant
\cite{nakanishi, gleeson}.

\textit{
It should be clear that the source of the problem lies in the 
fact that the set of two-particle null states that are supposed to be eliminated 
by the Lee-Wick real energy constraint is not Lorentz-invariant.  
This observation suggests that the problem may 
be fixable if we can recast the full theory in a manifestly covariant 
state space.   The discussion in the footnote also  suggests 
that, if we base our state space on holomorphic test functions,
the unconstrained theory may in fact have a \textit{hidden} covariance,
which can be made explicit by using the appropriate sets of 
generalized functions as continuum states. We now proceed to do so.}
 
In reference \cite{indef}, by the present author, a general
non-perturbatively well-defined 
Hamiltonian and path integral approach to indefinite inner
product theories was described.  The approach was
based on a class Gel'fand-Shilov distributions  on 
analytic test spaces that  provided
a very  natural description of  the 
complex-energy states that tend to arise in 
pseudo-unitary theories.

We will  continue that work and  formulate the current theory in a Gel'fand-Shilov
rigged state space based on 
a test space of  analytic functions. 
The space will be 
tailor-made for building covariant descriptions of complex mass 
degrees of freedom.
  There are many spaces
of analytic functions, but
our choice is constrained by the requirement that the test function space
be invariant under time evolution.  In the relativistic case, the relevant 
evolution (at least for the free theory) is given by the Klein-Gordon or
Dirac equations, and
we shall choose as our test space for a particle of mass $m$
 the Gel'fand-Shilov space of 
entire functions \cite{gelfand1, gelfand2, indef} on the mass shell hyperboloid 
parameterized by the  three-momentum $\bar k$ in some frame. 
$$
S_{1/2,A}^{1/2,B} (\bar k) \equiv \bigotimes_{i=1}^3 S_{1/2,A}^{1/2,B} (k_i),\qquad k_0 = \sqrt{m^2 + {\bar k}^2}, \quad \bar k = m \bar \lambda,\quad \lambda_i \in \mathbb R,
$$
The mass $m$ may be complex, in which case the momenta $\bar k$ are complex, as we discuss further below.
Members of $S_{1/2,A}^{1/2,B} (z)$ are entire functions with order of growth
$$
  \left|\phi(x + iy)\right|
    < C\, e^{-a|x|^2 + b|y|^2},  \qquad a = {1\over 2eA^2}, \qquad
    b = {eB^2\over 2}.
$$
Here $e$ denotes the base of the natural logarithm.
The parameters $A$ and $B$ will be chosen such that test functions will decay 
along the possibly complex directions 
$\bar k \in m {\mathbb R}^3$ for all masses 
$m$ in the theory.

The space $S_{1/2,A}^{1/2,B} (\bar k)$ is invariant under 
Klein-Gordon evolution \cite{gelfand2}, 
and has Fourier transform 
$$
  \mathcal F \left(S_{1/2,A}^{1/2,B}(\bar k)\right) = S_{1/2,B}^{1/2,A}(\bar x),
$$
where $\bar x$ parameterizes some space-like surface.  
Because the set of four-momenta
$$
  \left\{(k_0, \bar k)\,|\, k_0 = \sqrt{m^2 + {\bar k}^2}, \, \bar k = m \bar \lambda,\, \lambda_i \in \mathbb R \right\}
$$
is Lorentz-invariant for real or complex $m$, 
it follows that the test function space
 $S_{1/2,A}^{1/2,B} (\bar k)$ is in fact frame-independent.  
The full test space on an arbitrary number of
particles is generated via the Fock construction
from sums of tensor products of single-particle test functions.  We shall 
denote
the resulting Fock space by
$$
   \mathbb S_{1/2,A}^{1/2,B}.
$$

So far $A$ and $B$ have not been fixed.  However, let us make the 
physical assumption that the energy and each component of the 
momentum of all states
lie within a double wedge $|y| < \alpha |x|$ in the complex plane.  
By choosing $a>\alpha^2b$ in the above formula for the order of growth, 
or equivalently $AB < 1/e\alpha$, we can ensure that test functions 
decay as $e^{-\beta |z|^2}$  in any direction lying within the wedge. 

This choice of test space 
allows us to formulate complex mass-energy theories in a 
manifestly Lorentz-invariant way.  Consider, for example, the 
two-point function 
$$
  D(x - y) = \bra{0} \phi(x) \phi(y)\ket{0}
$$
of a scalar field with complex mass $M$.  A Lorentz-invariant
set of  one-particle states can be generated from the state
with four-momentum $(M, \bar 0)$ via Lorentz transformations 
to obtain the invariant hyperboloid
$$
  \left\{(k_0, \bar k)\,|\, k_0 = \sqrt{M^2 + {\bar k}^2}, \, \bar k = M \bar \lambda,\, \lambda_i \in \mathbb R \right\}.
$$
The momenta of these states are not real, so that the 
formal, manifestly Lorentz-invariant expression for the two-point function
$$
  D(x - y) = \int_{M\mathbb R^3} {d^3\bar p\over (2\pi)^3} \, 
   {e^{i p_\mu (x^\mu-y^\mu)}\over 2 \sqrt {M^2 + {\bar p}^2}}
$$
would naively appear to diverge due to the exponentially growing 
integrand.  However, the two-point function defines a perfectly meaningful 
distribution with respect to 
$S_{1/2,A}^{1/2,B} (\bar k)$.   Specifically, in momentum space
it is \textit{defined} via its effect on a test function $\phi$ as
$$
  \int_{M\mathbb R^3} {d^3\bar p\over (2\pi^3)} \, 
   {e^{i (p_0 x^0 - p_\mu y^\mu)}\over 2 \sqrt {M^2 + {\bar p}^2}}\, \phi(\bar p), \qquad p_0 = \sqrt {M^2 + {\bar p}^2},
$$
which converges due to the Gaussian decay of $\phi$ in the wedge regions
containing each component of the complex momentum in $M\mathbb R^3$.  
To obtain the  position space form of the distribution, 
note that we can deform the 
complex contours in the above convergent integral 
to the real line without encountering singularities to
obtain an integral over real momenta, and then carefully exchange 
limits as follows:
\begin{align}
  &\int_{M\mathbb R^3} {d^3\bar p\over (2\pi^3)} \, 
   {e^{i (p_0 x^0 - p_\mu y^\mu)}\over 2 \sqrt {M^2 + {\bar p}^2}}\, \phi(\bar p) \nonumber\\ 
  &\qquad =\int_{R^3} {d^3\bar p\over (2\pi^3)} \, 
   {e^{i (p_0 x^0 - p_\mu y^\mu)}\over 2 \sqrt {M^2 + {\bar p}^2}}\, \phi(\bar p)\nonumber \\
  &\qquad= \int_{R^3} {d^3\bar p\over (2\pi^3)} \, 
   {e^{- ip_\mu y^\mu}\over 2 \sqrt {M^2 + {\bar p}^2}}
  \int_{R^3} d^3 \bar x \, e^{ip_\mu x^\mu}\phi(x)\nonumber \\
  &\qquad= \lim_{\epsilon\to 0+}\int_{R^3} {d^3\bar p\over (2\pi^3)} \, 
   {e^{- ip_\mu y^\mu - \epsilon |\bar p|}\over 2 \sqrt {M^2 + {\bar p}^2}}
  \int_{R^3} d^3 \bar x \, e^{-ip_\mu x^\mu}\phi(x) \nonumber\\
  &\qquad= \int_{R^3} d^3 \bar x\,\phi(x)\left(\lim_{\epsilon\to 0+}\int_{R^3} {d^3\bar p\over (2\pi^3)} \, 
   {e^{ip_\mu\left(x^\mu- y^\mu\right) - \epsilon |\bar p|}\over 2 \sqrt {M^2 + {\bar p}^2}}\right). \label{ctoreal}
\end{align}
Note that the limit added in the third step is entirely superfluous, but allows us to exchange integrals in the last step.  The factor in parentheses can then be evaluated in terms of Hankel functions as in the case of real $M$.  The factor $e^{-\epsilon|p|}$ happens to be convenient for evaluating the integral 
for time-like $x - y$, and could have been replaced by something like $({\bar p}^2 + M^2)^{-\epsilon}$ for convenience in evaluating the integral for space-like $x - y$.

It is important to note that the use of an analytic test space does not
limit our ability to describe the world.  For example, any 
 square-integrable single-particle
wave function can be approximated arbitrarily closely by an element of 
$S_{1/2,A}^{1/2,B} (\bar k)$.  Furthermore, because the analytic 
test space is smaller
than the usual space of Schwartz test functions, the dual space of 
generalized functions is \textit{larger} than the space of Schwartz 
distributions.  We therefore gain the ability to describe extra interesting 
generalized states, such as states with complex energy-momentum, that 
are not Schwartz distributions.  Indeed, there is
no first principle that forces one to accept Schwartz distributions 
as canonical,
and in fact Schwartz distributions are known to be insufficiently
general for describing such useful entities in ordinary scattering theory
as the complex energy-momentum Gamow states describing resonances
or unstable particles
\cite{indef, berggren, newton, buchleitner, moiseyev, hagen, ho, madrid}.

One important consequence of choosing an analytic test space is that the 
continuum part of the spectral decomposition of operators is no longer 
unique.  
This gives us considerable freedom in choosing sets of continuum states to
use  in 
completeness relations.
In the present case, this freedom will allow us to find an equivalent formulation 
of the full theory based on a Lorentz-invariant state space.  

As a simple one-dimensional example of this freedom, consider an element 
$\phi \in S_{1/2,A}^{1/2,B}(x)$.  We have 
\begin{align*}
  \phi(x) = \int_\mathbf{R} {dk\over 2\pi}\, e^{ikx}\,\tilde \phi(k),
\end{align*}
where the Fourier transform $\tilde \phi \in S_{1/2,B}^{1/2,A}(k)$ is also an entire 
function.  As a result, we can deform, say, a finite section of the integration 
path from the real line into the complex plane.  Call the deformed path $\Gamma$.  
Then 
\begin{align*}
  \phi(x) &= \int_\Gamma {dk\over 2\pi}\, e^{ikx}\,\tilde \phi(k) \\
         &= \int_\Gamma {dk\over 2\pi}\, e^{ikx}\int_\mathbf{R} dy\, e^{-iky}\,\phi(y).
\end{align*}
The integral over $y$ converges for any complex $k$ due to the Gaussian decay 
of $\phi \in S_{1/2,A}^{1/2,B}(x)$.  But for complex $k$, we have 
$$
e^{-iky} = (e^{ik^*y})^* = {\braket y {k^*}}^* = \braket {k^*} y.
$$
Therefore 
$$
  \tilde\phi(k) = \braket{k^*} \phi,
$$
and
\begin{align*}
  \phi(x) &= \int_\Gamma {dk\over 2\pi}\, \braket{x}k \braket{k^*}\phi.
\end{align*}
In other words, the set of complex-momentum null generalized 
states lying along
$\Gamma$ is also complete when acting on 
test functions.  Formally, the identity on $S_{1/2,A}^{1/2,B}(x)$
has many possible spectral decompositions in terms of momentum eigenstates,
such as
\begin{align*}
  \phi(x) &= \int_\mathbf{R} {dk\over 2\pi}\,\ket k \bra k \\
   &= \int_\Gamma {dk\over 2\pi}\, \ket k \bra{k^*} \\
      &=\half\int_\Gamma {dk\over 2\pi}\, \ket k \bra{k^*}
             + \half\int_{\Gamma^*} {dk\over 2\pi}\, \ket k \bra{k^*} \\
    &= \cdots .
\end{align*}
All these representations are equivalent for doing physics,
 given the assumption that we can
approximate
physical wave packets by test functions.  However, notice that the second
completeness relation does not extend to the formal generalized states $\ket{k}$
and $\ket{k^*}$ themselves since, for example, $\braket{k^*}{k^*} = 0$.  
In other words, we cannot carelessly  generalize 
relations outside the domain of the distributions involved.\footnote{
Although the third line above can be made to works also when applied 
to  generalized momentum states as long as 
we choose the inner products $\braket k {k^*}$ proportional to the appropriate 
delta functions with respect to the integrals along $\Gamma$ and $\Gamma^*$.}

This is analogous to what is happening in our field theory.  Consider the 
previously described 
scattering of a non-ghost incoming wave packet in $\mathbb S_{1/2,A}^{1/2,B}$.
We could deform the integration contour lying 
above the fish-shaped non-analyticity region to the real line along the analytic
continuation of the integrand in the amplitude.  
The analytic continuation was Lorentz-invariant
and had a cut along the real axis starting at $s = (M + M^*)^2$ 
\cite{lee2, nakanishi, gleeson}.  
This suggests  that the amplitude obtained from 
the set of 
intermediate complex $s$
two-ghost states indexed by the points in the non-invariant fish-shaped region 
is equivalent, modulo the $\mathbb S_{1/2,A}^{1/2,B}$ test space,
to the amplitude obtained from some set of real $s$ intermediate states
indexed by the points along the cut on the real axis.  
As we shall see, the latter set of states can be chosen to be
manifestly Lorentz-invariant, and so that their momenta 
and energies are 
real.

In the unconstrained theory, 
it therefore will not make a difference, for 
the calculation of scattering amplitudes of non-ghost states, whether
we use this manifestly Lorentz invariant set of intermediate real-energy 
ghost states or the Lorentz
non-invariant fish-shaped set of complex-energy ghost states.
Either set of continuum states can be expressed in terms 
of the other.  With one set, Lorentz invariance is explicit;
with the other, it is hidden.  
However, the choice 
will make a difference when applying the Lee-Wick real energy constraint,
which we will do 
in a non-perturbative setting where 
we apply  an infrared regularization (which discretizes momenta
and causes the continuum states to become actual normalizable states),
then  apply the Lee-Wick real energy constraint, and then remove the
regularization.  This procedure will give a Lorentz-invariant result 
only if we use the manifestly Lorentz-invariant multi-ghost states.

\textit{
We will therefore insist that any non-perturbative approach to
applying the Lee-Wick real energy constraint 
should be based, from the start, on a 
manifestly Lorentz-invariant state space.  }

We now proceed to  find a Lorentz-invariant set of
two-particle states that can represent  a ghost of complex
mass $M$ together with the corresponding anti-ghost of   mass
$M^*$.  An obvious, but unworkable,  approach would consist of
following the Fock construction by making
two-particle states from the free 
tensor product of the invariant mass hyperboloids, discussed above  for
a single particle, of masses $M$ and $M^*$.
This space, convenient as it would have been, is unsuitable for the 
following reason.  It is not difficult to see that 
a given component of the combined space-like momentum of  the
two particles in this parameterization can be 
an arbitrary complex number.  As a result, the scattering 
amplitude with intermediate two-particle states in this space
will include an integration 
over the complex plane, weighed by the incoming wave packet in 
 $\mathbb S_{1/2,A}^{1/2,B}$.  However, as we have specified the 
test space, this wave packet 
has exponential growth in the imaginary direction, and we
will not get a finite result.  

Therefore, we have to be more careful in specifying an invariant 
two-ghost state space.  
We will first write down one particular suitable choice of invariant space
and then justify it below by considering the explicit expression for 
certain  Hamiltonian matrix elements.  

In (\ref{ctoreal}) we saw that single-particle 
three-momenta can be deformed from complex to real modulo the analytic test
space.   
Let $b_{\bar k}$ and $\tilde b_{\bar k}$
be creation operators for the single ghost states with four-momenta
$(\sqrt {M^2 + \bar k^2} , \bar k)$ and  $(\sqrt {(M^*)^2 + \bar k^2} , \bar k)$
respectively, where $\bar k$ is real.  The particular 
 Lorentz-invariant space of
 real-energy two-ghost states that we will use is 
spanned by 
\begin{align}
\left\{\mathcal{L}\left(b_{\bar k}\tilde b_{-\bar k}\right) \ket{0}
|\, \bar k \textrm{ real}, \mathcal{L} \textrm{ a Lorentz transformation}\right\}.  \label{2partinv}
\end{align}
Here
$$
  \mathcal{L}\left(b_{\bar k}\tilde b_{-\bar k}\right)
   \equiv \mathcal{L}\left(b_{\bar k}\right)
    \mathcal{L}\left(\tilde b_{-\bar k}\right),
$$
and, taking $\mathcal{L}$ to be, for example, a boost in the $x_3$-direction, we
have 
$$
  \mathcal{L}\left(b_{\bar k}\right) \equiv 
    b_{(k_1, k_2, \sqrt{M^2 + k^2} \sinh \beta + k_3\cosh \beta)}
$$
and likewise for $\tilde b_{\bar k}$ with $M$ replaced by $M^*$.  
The transformed operators individually have complex momenta, but it is 
easily checked that the combined four-momentum
 of the two-particle state is real, and therefore fall inside the 
wedge where the test functions have exponential decay.
This space of states should  therefore be suitable for the 
calculation of scattering amplitudes with incoming wave packets
in the space $\mathbb S_{1/2,A}^{1/2,B}$. 
\textit{It should be noted that this space of two-particle states is not a free tensor 
product of single ghost states.}\footnote{
There are other possible choices of Lorentz-invariant two-ghost state
spaces.  For example, we could have taken the space-like momenta of both
particles from $M\mathbb R^3$ instead of $\mathbb R^3$.  We conjecture that 
all such choices should be equivalent as far as scattering of real-energy 
non-ghost particles are concerned.}  

To justify this choice of two-particle space 
from a  non-perturbative point of view, we need to find the matrix 
elements of the Hamiltonian for 
creation of the two-ghost states in this invariant set
from a single-particle state.   (More generally, we need the $m\to n$
particle matrix elements, which can be obtained via a similar
procedure.)
Assuming for simplicity that all fields are scalars, a typical 
cubic term in the Hamiltonian will have the form 
\begin{equation}
  \int_{\mathbb R^3} {d^3 \bar p\over (2\pi)^3}\, {1\over 2E^m_{\bar p}}\,
   a_p \int_{\mathbb R^3} {d^3 \bar k\over (2\pi)^3}\, 
    {1\over 2E^M_{\half\,\bar p + \bar k}}\,{1\over 2E^{M^*}_{\half\,\bar p - \bar k}}\,
     b^\dagger_{\half\,\bar p + \bar k} \tilde b^\dagger_{\half\,\bar p - \bar k}
\label{originalH}
\end{equation}
with respect to invariantly normalized real-momentum creation-annihilation 
operators.  Here $a$ is the ordinary scalar of real mass $m$ and 
$b$ and $\tilde b$ the ghosts of complex masses $M$ and $M^*$ respectively.
Acting on one-particle states, this term will not explicitly 
create two-ghost states in
the manifestly covariant space constructed above, but modulo the analytic test 
space, we can deform the
integral over $\bar k$ to do so.  Without loss of generality,
take 
$$
  \bar p = (0, 0, p_3) \equiv (0, 0, \sinh\beta 
    \left(\sqrt {\bar k^2 + M^2} + \sqrt {\bar k^2 + {M^*}^2}\right)),
$$
which defines $\beta$, and deform 
the integral over $\bar k$ to
\begin{equation}
\int_{\mathbb R} {dk'_1\over (2\pi)}\int_{\mathbb R} {dk'_2\over (2\pi)}
 \int_{\Gamma_{k'_1, k'_2}}{d k'_3\over (2\pi)}\, 
    {1\over 2E^M_{\half\,\bar p + \bar k'}}\,{1\over 2E^{M^*}_{\half\,\bar p - \bar k'}}\,
     b^\dagger_{\half\,\bar p + \bar k'} \tilde b^\dagger_{\half\,\bar p - \bar k'}
\label{deformedH}
\end{equation}
where the complex curve $\Gamma_{k'_1, k'_2}$ in the $k'_3$ plane 
is defined by the equation
\begin{align*}
   k'_3 &= \half \sinh \beta \left( \sqrt {\bar k^2 + M^2} - \sqrt {\bar k^2 + {M^*}^2}\right) + k_3 \cosh \beta, \\
 &\qquad k_{1,2} = k'_{1,2},\quad
   k_3 \in \mathbb R.
\end{align*}
This particular deformation is justified when we take the matrix elements
of the expression between the one-particle and two-ghost subspaces of 
the analytic test space
$\mathbb S_{1/2, A}^{1/2, B}$, and different deformations will be 
needed for higher matrix elements.  The curve $\Gamma_{k'_1, k'_2}$ 
asymptotically 
approaches the real line as $k_3 \to \infty$, 
and therefore lies within the region of Gaussian decay of the
test functions.  Since 
\begin{align*}
  \half\, \bar p + \bar k'
   &= (k_1, k_2, \sqrt{\bar k^2 + M^2}  \,\sinh \beta + k_3 \cosh \beta), \\
 \half\, \bar p - \bar k'
   &= (-k_1, -k_2, \sqrt{\bar k^2 + {M^*}^2}  \,\sinh \beta - k_3 \cosh \beta),
\end{align*}
it follows that  the states $b^\dagger_{\half\,\bar p + \bar k'} \tilde b^\dagger_{\half\,\bar p - \bar k'}\ket 0$ lie in the Lorentz invariant two-ghost 
state space, and 
we can read off the matrix elements linking the one-particle states to the 
manifestly invariant two-ghost space from the deformed Hamiltonian.   

Note that this deformation is only valid for the specific $1\to 2$ matrix elements considered above.  
For general $m\to n$ matrix elements, Lorentz-invariant state spaces
of multi-particle states 
can be generated in a similar way as for the two-ghost states above.  
However, different deformations of the cubic term in the Hamiltonian will
generally be 
required to obtain the matrix elements linking these invariant 
spaces.  A fuller analysis is 
left for future work.

Given the equivalence, modulo the test space,
 of the original and the deformed Hamiltonian
matrix elements, it follows that both should 
give equivalent amplitudes for scattering of wave packets of
non-ghost states.  It is the 
amplitude calculated using the original 
Hamiltonian matrix elements (\ref{originalH}) 
that has the previously mentioned non-Lorentz-invariant 
fish-shaped non-analyticity region in $s$-space (corresponding to the 
complex energies of the intermediate ghost states) with the 
contour of integration running \textit{above} 
this region, and it is the
 amplitude calculated using the equivalent deformed matrix
elements (\ref{deformedH})  that
has the Lorentz-invariant cut in $s$-space
 along the real line corresponding to the 
real energies of the intermediate two-ghost states from the 
Lorentz-invariant set.  The equivalence of these amplitudes
means that the the integrand in the latter 
 should be the continuation 
of the same analytic function as that in the former 
 from outside the fish-shaped region.
This is supported by the observation that 
the continuation of the former amplitude from outside the 
fish-shaped non-analyticity region does give
 an analytic function with a
cut starting at $(M+M^*)^2$ along the real axis in the $s$-plane, whose
explicit form can be found in \cite{lee2, gleeson}.

\textit{We have succeeded in rewriting the amplitudes in terms
of a Lorentz-invariant set of two-ghost states.  However, at this
point it appears that these 
 states have real energy, and therefore  will 
not be eliminated directly by the Lee-Wick real energy constraint.  
This seems, at first glance,
to be a disaster for unitarity.  However, we shall indicate that the problematic 
negative-definite states do get eliminated after taking into account their 
interactions in a non-perturbatively well-defined way.  We shall further
see that, 
for the diagram 
we consider, the method
will effectively reproduce Lee's and Cutcosky et al.'s\ prescription for projecting out these states
\cite {lee2},
without relying on their ad hoc assumptions.}

\section{Nonperturbative unitarity via  elimination of  multi-ghost states}

Our analysis so far would suggest that the theory contains 
real-energy multi-ghost states.  If this were the case, there would be 
a problem with unitarity, since the theory would 
have negative-metric states that could not be eliminated by 
the Lee-Wick real energy constraint.  In this section, we will argue
that, in a specific non-perturbatively well-defined approach,
 the energies of these states in fact become non-real
due to interactions.  As a result, many multi-ghost states
 will be 
eliminated by the Lee-Wick real-energy constraint.
The rest will be removed by a ghost number constraint on 
the asymptotic states.\footnote{
In the next section, we will describe an alternative,
less general but very simple method,  that may also, in certain cases, be useful
for eliminating these multi-ghost states.  The method in that section 
constrains the mass of the ghosts to be of the same order as the
total mass-energy of the modeled universe.  The more general method 
in the current section imposes no such constraint on the ghost mass.}

The simplest case where the problem
might occur are the two-particle states consisting of 
a ghost and anti-ghost degree of freedom 
of the original Lagrangian.  We have seen that the mass of these
particles remains real in the presence of interactions, and so the 
two-particle states would be expected to have real energies.
A slightly more complicated case where this may occur
are the two-particle states consisting of a ghost positronium particle
and anti-particle.  These have complex conjugate masses, but as we saw
in the previous section, we can re-express the two-particle states
in terms of a Lorentz-invariant space of real-energy states.\footnote{
In this case, individual particles have complex energy, and are therefore 
null states.   However, the degenerate real-energy 
(anti-)symmetric combinations
of these are 
positive (negative) definite.}
Since in both cases we find negative metric  states with  real energy, this seems to be a 
disaster for unitarity, since the Lee-Wick real energy constraint will not eliminate such states.

\textit{However, interactions  can be expected
 quite generally to move the mass-energy of these 
states away from the real axis into the complex plane, at which point the 
Lee-Wick real energy constraint can be applied to eliminate them.}

Note that this statement would \textit{not} be true for ordinary 
unitary local field theories where, given the exact
single-particle energies, we can  infer
 the exact energies of two-particle
states to be simple sums of single-particle energies
by simply increasing the distance between the two particles
to the point where the interactions are 
sufficiently negligible.  
For interactions to make a difference to the two-particle
energies, we could impose a large-distance (infrared) cutoff,
which would prevent us from increasing the separation 
arbitrarily.  However, in ordinary  unitary local field theories, the 
energy shift due to interactions is real and vanishes once
we remove the cutoff.

In pseudo-unitary theories, the energies of 
any two-particle negative-metric states will likewise be 
modified by interactions when we impose a large-distance
cutoff preventing indefinite separation.  However, as we 
will argue, with certain mild conditions,
these modified 
energies will generally  become non-real.  
Therefore, as long as we implement the Lee-Wick real energy constraint
\textit{before}  removing the large-distance cutoff, these
states will be eliminated.

Technically, 
what happens is analogous to our previous discussion of the single pole due to a 
negative-metric state lying on a cut due to positive definite states.  
Unless a symmetry forbade the interaction, the negative-metric state    
became unstable with respect to decay into the 
positive definite states forming the cut,   and as a result
the pole split  into a pair of complex conjugate positions on the 
physical sheet away from the real 
axis.
The single negative metric state can linearly combine with positive metric
states to give precisely
two dual null states.  All other states had to remain on the real
axis.  

In general, the two-ghost negative metric cut 
is degenerate with a cut due to positive-metric  states of much lower mass, and 
to which the two-ghost states can transition.
For example, a pair of complex ghost-positronium atoms can annihilate and 
transition to a multi-photon or multi-lepton
state.  We then assume a long-distance  regularization   that 
temporarily discretizes the four-momenta.  Because the masses  of the 
constituents of the
negative metric two-ghost states are by assumption
 much larger than those of the 
constituents of the positive-metric
transition products,
as we increase the large distance cutoff the positive-metric states can, 
at the high energies where the two-ghost 
states appear, be 
represented to an increasingly  good approximation by a continuous cut
in $s$, while the two-ghost
states, which have comparatively much smaller phase volume
in the now discretized momentum space, may
still be considered discrete.  Just as in the single-pole case, the
interaction will split each 
negative-definite state into a pair of dual null states with complex-conjugate 
mass-energies.  

The conditions for this argument to work can be expressed in
another way.  As we saw by way of the toy  example in section
\ref{sectionmass}, interactions
between a positive metric state and a negative metric state, both
of whose unperturbed
energies are real, will generally
split the spectrum into a pair of null states with complex conjugate energies
as long as the distance between the unperturbed energies is 
comparable to the scale of the interaction.  In the  field theory
case, as long as the ghost masses are much larger than
the masses of the transition products (in this case 
lepton or photon multi-particle states), each two-ghost state will, 
as we increase the large-distance cutoff, 
approach  degeneracy with many positive-metric states into which they can
transition.  Thus, the conditions are met for all these negative-metric states
to be removed from the real spectrum by the interaction, 
combining with suitable positive metric states to form dual
null states with complex conjugate energies.  Only positive-metric
states should remain in the real spectrum.  

According to our introductory arguments, 
as we remove the long-distance cutoff,  the energies of these null 
states will become real.  For this reason, we propose imposing 
the Lee-Wick real energy constraint before removing the cutoff.  In particular,
the procedure is as follows:

\textit{Impose a large-distance cutoff 
and show that the
 interaction moves negative-definite 
states away from 
the real line.  Then apply the Lee-Wick real energy constraint before removing 
the cutoff.  In symbols, the state space is obtained as follows.}
$$
  \lim_{\textrm{L}\to\infty} \textrm{P}_\mathbb{R}^{H_L} \mathcal H,
$$
where $\mathcal H$ denotes the full indefinite inner product state space
and  $\textrm{P}_\mathbb{R}^{H_L}$ is the projection onto the 
real-energy eigenspace of the regularized Hamiltonian $H_L$, where 
$L$ is the long-distance cutoff.  

It is interesting that this method will, in a simple example, produce a
result very much like  Lee's 
ad hoc prescription (and that of Cutkosky et al.)
for eliminating these continuum states.  Our method provides 
a justification for their prescription.   Unlike their approach, 
ours is non-perturbatively well-defined in principle, and 
we rely on higher-order interaction 
effects.

As an illustration, 
let us reconsider a scattering process in which the photon appears as 
intermediate state in the $s$-channel.  
The transverse part of the photon propagator is
\begin{align}
  D(q^2) = {-ie^2\over q^2}\,{1\over 1 + e^2 \,\Pi(q^2)}, \label{photonprop}
\end{align}
where the vacuum polarization contribution was
calculated in section \ref{sectionsafe} to be
\begin{align}
  \Pi (q^2) &=  {e_\Lambda^2\over 2\pi^2}
   \int_0^1 d\beta \, \beta(1-\beta)\,\ln{{M^2 - \beta(1-\beta)\,q^2 - i0}\over{m^2 - \beta(1-\beta)\,q^2 - i0}}.
\end{align}
The scattering amplitude is then proportional to 
\begin{align}
  D(s) = -ie^2\left({1\over s} + \int_{4m^2}^\infty ds'\,{\rho_m(s')\over s -s' + i0} + \int_{4M^2}^\infty ds'\,{\rho_M(s')\over s -s' + i0}\right).
\label{photondisp}
\end{align}
Here $\rho_m$ represents the unitarity sum over intermediate 
pairs of ordinary fermions, and $\rho_M$ the unitarity sum over intermediate pairs
of ghosts.   

This expression is valid as as it stands below the 
$2M$-threshold.  However, above the threshold, the off-shell amplitude will be 
modified when we apply 
the non-perturbatively well-defined projection described above.
This will be fortunate, because the cut starting at the $2M$ threshold is 
incompatible with unitarity, corresponding as it does to the production of 
negative-definite states.   

We now follow our suggested procedure and impose a large-distance
cutoff, which discretizes the momenta.  The cut
starting at $4M^2$ will become a series of poles.  The ghost states represented by 
these poles can annihilate to give a virtual photon that decays to ordinary 
electron-positron states of much lower mass.  According to the general
 argument above, this interaction will make
 these two-ghost poles move away from the real 
axis to give a pair of complex conjugate poles corresponding to null states.  
Only positive definite states remain on the real line.  

The Lee-Wick real energy constraint, which eliminates the null states, 
is equivalent to changing the integration contour in the $s$-plane,
which in the full theory runs above all these complex 
poles, to run along the real axis. 
Our non-perturbatively well-defined prescription consists of doing  this change of contour 
before removing the regularization.    
Therefore, half of the null state poles will unambiguously
lie at complex positions above the integration contour and
half below.  The result when we do finally take the continuum limit
is therefore equivalent to calculating the amplitude with the discrete
two-ghost poles becoming a pair of 
cuts, one running infinitesimally above the real axis
and the other running below it, and with the integration contour running along 
the real axis between them.  The cuts meet at the threshold $2M$,
so that the values of the amplitude below and above the threshold belong to separate 
sheets of an analytic function.

Interestingly, the numerical value of the amplitude above the threshold can 
in fact be read off from the original continuum calculation of the amplitude.
First we note that the photon propagator was obtained from a
geometric series that already
included interactions between the negative
 and positive definite 
states that should be 
 responsible for moving the ghost pairs into the complex plane.  
Indeed, the propagator summed sequences of bubbles to all orders, including 
diagrams in which ghost pairs and ordinary fermion pairs alternate.  
Therefore, in the no-cutoff
limit, the result (\ref{photondisp})  already takes into account
the contribution of the conjugate pair of cuts described in the previous 
paragraph.  As far as the 
integration contour of the full unconstrained theory,
 which runs above all singularities in the 
$s$-plane,
is concerned, these cuts can be deformed to the real line and the result is
indistinguishable
from the single cut in the expression (\ref{photondisp}).  However, according to 
the reasoning of the previous paragraph, the integration path in the 
projected theory must go between the conjugate cuts
infinitesimally above and below 
the real line.  The one function  that will give the correct result with 
respect to both the unconstrained  theory's contour above the singularities
and the constrained  theory's 
real-line contour can therefore only be
\begin{align}
  D(s) &= -ie^2\biggl({1\over s} + \int_{4m^2}^\infty ds'\,{\rho_m(s')\over s -s' + i0} \\
  &\qquad \qquad \qquad\ + \half\int_{4M^2}^\infty ds'\,\rho_M(s')\biggl\{{1\over s -s' + i0} + {1\over s -s' - i0}\biggr\}\biggr) \\
&= -ie^2\left({1\over s} + \int_{4m^2}^\infty ds'\,{\rho_m(s')\over s -s' + i0} + \mathrm{P}\int_{4M^2}^\infty ds'\,{\rho_M(s')\over s -s'}\right). 
\end{align} 
\textit{This result is now consistent with unitarity, since the last term in 
parentheses does not contribute an imaginary part for real $s$.}

Similar considerations will apply to states of higher ghost number.
The above construction (\ref{2partinv}) of a Lorentz-invariant space of
two-ghost states can be generalized to obtain Lorentz-invariant state
spaces containing an arbitrary number of ghost excitations.  
For example,
take all individual space-like momenta real in the center of mass frame, 
and then generate an invariant set of states by applying the Lorentz group.
Other Lorentz-invariant constructions are possible (as we saw before, 
there is no unique continuum basis dual to spaces of holomorphic
test functions), but this non-uniqueness  should not matter for 
calculating amplitudes of non-ghost particles, which should be unique.  
Again, these states 
are not free tensor products of single-particle states.  However, 
their matrix elements in the Hamiltonian can again be obtained
via a similar procedure as was used in the previous section to obtain matrix 
elements involving two-ghost states.
A more exhaustive analysis is deferred to future work.

Again, this construction will in general
 give negative-metric ghost states of 
real energy, which would be disastrous for unitarity
unless we can eliminate them by applying the Lee-Wick real energy constraint
as we did to eliminate the two-ghost states above.  
The construction will also give positive-metric states containing ghosts 
(for example, any state containing any number of 
ghosts or an even number of 
anti-ghosts).  
Of these, our non-perturbatively well-defined limiting procedure should eliminate all
multi-particle states containing at least one ghost and anti-ghost  
(irrespective
of whether the overall metric of the state is positive or negative)
since these can annihilate pairwise and transition
 into 
multi-photon or multi-lepton  states
 consisting of particles of much lower mass.
The metric of 
the resulting
state 
has sign opposite to the original.
As before, for large enough distance cutoff,
these ghost states appear discrete compared to the effective continua 
of ordinary physical particles into which they can transition.
Thus, the conditions are satisfied
for the state to split into a pair of complex-conjugate energy
states.   Again, applying 
the Lee-Wick real energy 
constraint and then removing the cutoff  will eliminate these
states.  

\textit{
Our method so far has eliminated all states 
containing at least one ghost and one antighost.  We now 
discuss the  elimination of
 the remaining states that contain ghosts.}

We now argue that we will obtain a unitary ghost-free theory
 by further restricting the space of asymptotic  scattering states
(in other words, the eigenstates of the free Hamiltonian
$H_0$ used to set up the scattering problem in section \ref{scattering})
to the zero ghost number sector. Ordinarily, the interaction
would allow ghosts and anti-ghosts to be created in pairs
from initial  states with zero ghost number, 
but since all states containing such pairs are already removed from  the 
asymptotic spectrum by the Lee-Wick real-energy constraint, 
no ghosts will be created by 
the S-matrix 
of the constrained theory.\footnote{However, note that 
the exact states that are obtained from zero-ghost asymptotic
states via the mappings $W_{\pm}$ in section \ref{scattering}  are still in general 
nontrivial superpositions containing ghost terms.  The zero-ghost constraint 
is imposed on the eigenstates of $H_0$ labeling the 
asymptotic states and make sense only in the context of the S-matrix.}
\footnote{
The zero ghost number constraint also 
eliminates otherwise stable states  states containing only ghosts
 or only anti-ghosts, which  have real mass-energy even in the presence of 
interactions, as we saw in section \ref{sectionmass}.  }

\textit{
To summarize, we have argued that
we can obtain a unitary S-matrix
applying the Lee-Wick real energy constraint and large-distance regularization
in proper order, and by restricting the asymptotic state space to the 
ghost number zero sector.  }

\section{A second mechanism for unitarity}
\label{sectionsecond}

In the previous section, we described a general method for 
eliminating real-energy multi-ghost states in a non-perturbatively well-defined
framework that consists of  applying
the usual Lee-Wick real energy constraint before removing the long-distance
cutoff. 

In this section, we will describe a simpler but less general  alternative
construction that may be useful 
in specific cases.   However, the approach of this section requires 
the ghost mass to be of the order of the mass-energy of the modeled 
universe. 

Our approach here is based on the idea that  a theory describing 
a universe of finite mass-energy only needs to be unitary 
below this scale.  Negative-metric states with mass-energy 
larger than the total mass of the universe are never inhabited, 
so a  theory that has such states can still be unitary for all  
physically realizable  states.  For the sake of this argument,
we assume a flat Minkowski universe, and proceed as follows:
\begin{itemize}
\item
Assume  that the mass-energy of
the universe is bounded by a finite number.
More precisely, assume that the 
state of the universe is expressible as a superposition of eigenstates 
of the Hamiltonian, the energies of which have an upper bound. 
Any separable 
subsystem of the universe will 
necessarily have a mass-energy lower than this number.
\item
A quantum theory of the universe must be able to predict
a unitary S-matrix for the 
universe and for any separable subsystem of the universe.
 A quantum theory can be a complete description of
 a universe 
of bounded mass-energy without needing 
to model states of mass-energy above this bound
unitarily, or at all.  
\item
We can retain  exact Lorentz invariance, as well as 
locality and cluster decomposition to within the 
observational limits implied by these assumptions, 
by using pseudo-unitary quantum field theories that are 
unitary at mass-energies lower than the universe bound.   
\end{itemize}

This approach immediately considerably widens the range of possible
quantum field theories that may be considered as complete theories 
describing universes of bounded mass-energy.
In particular:

\begin{quotation}
\noindent
{Consider an indefinite metric quantum field theory that is asymptotically safe,
so that it is ultraviolet compete and has exact Lorentz invariance.
Suppose that the 
center of mass energy of the lowest negative definite eigenstate with 
real energy of the 
Hamiltonian $H$ is $M$.  Then the subspace of real-energy
eigenstates of $H$  satisfying the constraint 
$P_\mu P^\mu < M^2$ is positive definite.  
By restricting to this subspace, we obtain a unitary and Lorentz-invariant 
quantum theory.}
\end{quotation}
Note the following:
\begin{itemize}
\item
The requirement of asymptotic safety is very important.  It is a 
necessary condition for theory to have a 
well-defined continuum limit, without which Lorentz-invariance would not
hold, and the constraint on the state space would not be invariant.
\item
The resulting theory has exact Lorentz covariance.  Its space of physical states
is Lorentz-invariant.
\item
Because the physical sector is embedded in a larger quantum field theory, 
and we expect cluster decomposition to hold in the latter due to the local
nature of its formulation, independently of the signature of inner product, we also
expect to see cluster decomposition in the physical sector to within the
observational limits implied by the constraint on physical states.
\item
It is important to note that virtual momenta in Feynman loops are \textbf{not} 
cut off.  Loops are already finite in the unconstrained theory 
through asymptotic cancellations as
in section \ref{sectionsafe}, or through renormalization as in section 
\ref{sectionmass},
before imposing the constraint.
The constraint $P_\mu P^\mu < M^2$ selects a
subspace of physical states 
post hoc from a larger indefinite metric theory that is well-defined
at arbitrarily high energy, but does not affect virtual loops.  
\item 
Since the theory is ultraviolet complete,
he constraint is not a regularization, 
and it is not a measure of our imperfect knowledge of higher-energy
physics.  
\end{itemize}
The constraint on $P_\mu P^\mu$ above implies the Lee-Wick 
real-energy constraint.  
As far as S-matrix calculations are concerned,  we have to use
the Lee-Wick methods as before to eliminate the exponentially
growing contributions of any complex-energy eigenstates
the theory may have.  Imposing  the $P_\mu P^\mu$ constraint on the free 
asymptotic states in the far past will ensure, as in ordinary field
theories, that no states with real mass-energy above the bound
will be produced.  However, as we saw before, any 
non-zero overlap of  free states 
 with complex-energy eigenstates
of the full Hamiltonian leads to exponential
terms that cannot be ignored.  

Stated another way, 
applying the constraint $P_\mu P^\mu$ on the asymptotic scattering states
is complicated by the fact that, as we have seen, we are 
not entitled to regard scattering states in the far past or future
as approaching 
eigenstates of the free Hamiltonian, even though to a local 
observer they become locally indistinguishable from free eigenstates 
to arbitrary accuracy.  Small 
discrepancies
due to overlaps with complex-energy states  grow 
exponentially under time evolution.  
Constraining the spectrum to real energies 
can be done by identifying and discarding
these  exponential contributions, just as we discussed before.

 The feasibility of a constraint of this type 
in modeling somewhat realistic universes is 
supported by the calculation in section \ref{sectionsafe}.
In particular, taking the fine structure constant 
$$
  \alpha \sim {1\over 137}
$$
in the theory of ghost QED, and 
$m \sim 0.511\,\textrm{MeV}$ and,
as using, for the sake of applying the 
constraint   $P_\mu P^\mu < M^2$, a ghost 
 mass $M$ as large as the mass-energy of the
visible part of our universe,
estimated as
$$
M \sim 1.25\times 10^{82}\, \textrm{MeV},
$$
 we  find a small bare coupling 
$$
  \alpha_\infty \sim {1\over 100},
$$
and, more importantly,  an effective perturbative expansion parameter 
to first order
\begin{align}
  {\alpha_\infty \over 3\pi}\, \ln{M^2\over m^2} \sim 0.42  < 1,
\label{firstorder2}
\end{align}
despite the enormous value of $M/m$.
In other  words, perturbation theory remains valid for all 
energies, and the theory could have been 
a complete theory if the universe consisted of 
QED only.  In the approach of this section based on 
the $P_\mu P^\mu$ constraint, and given the measured values of 
$\alpha$ and $m$, ghost QED can only 
be a complete  theory if the mass-energy of 
the modeled universe is finite and less than $M$,
where $M$ is constrained by the requirement that 
(\ref{firstorder2}) be sufficiently smaller than unity.  

In the mechanism for obtaining unitarity discussed in the 
previous sections, the same upper limit on $M$ is valid.
However, in that case, $M$ did not limit the mass-energy 
of the universe, and could in fact be much smaller than the 
mass-energy  of the universe.  
In that case, $M$ is  the scale at which 
non-local or unusual causal effects become observable,
as we shall see below in section \ref{sectioncausality}.

\section{Hierarchy and naturalness}

In our second approach to unitarity, based on the constraint
$P_\mu P^\mu < M^2$  discussed in the previous section, there 
is an amusing naturalness 
relationship between the fine-structure constant
and the matter-ghost hierarchy constrained by the 
the mass-energy of the QED universe.   

Specifically, the value 
of the fine structure 
constant may be related to the upper bound on the mass-energy of the universe
via a naturalness argument.  

As we saw, for the theory to be 
well-defined, the leading loop expansion parameter
\begin{align}
  {\alpha_\infty \over 3\pi}\, \ln{M^2\over m^2}, 
\label{firstorder3}
\end{align}
which is the largest perturbative parameter in the
theory,
must be smaller than one.  If we consider
it natural that this parameter be of order  unity, we 
may conclude  that 
the  fine-structure constant must  be small  if 
 the mass-energy of the universe is 
large compared to the 
mass of the electron.
 In other
words, an observation of the total mass-energy of the QED
 universe can be used to 
infer, via naturalness, an upper bound for the value of the fine structure constant.

As we saw in the previous section, the value of $0.42$ for the loop 
expansion parameter, which 
is very much of order unity and presumably ``natural'', 
 is remarkably compatible with  
realistic values
of $\alpha$, electron mass, and (visible)  universe mass energy in our own 
universe.  Thus, the large  mass of the universe may be invoked 
to ``explain'' the 
small observed value of the fine structure constant.

The argument so far assumed 
our second approach to unitarity based on the $P_\mu P^\mu$
constraint.  In the first approach, the ghost mass $M$ did not 
constrain the mass-energy of the modeled universe.  
In this case, we can still make a slightly weaker statement 
based on naturalness.  Specifically, the argument remains true
that it is natural for a larger hierarchy $M/m$ to be associated with 
a smaller  fine structure constant
$\alpha$, with the
 logarithmic relationship implied by (\ref{firstorder3}).
The difference in this case is that now   $M/m$ is unrelated to 
universe size.

\section{Causality}
\label{sectioncausality}

An interesting feature of the unitary theories  obtained by
the Lee-Wick real energy constraint  is an unusual causal behaviour.
Aspects of this were described by Lee and Wick
and further discussed by Coleman \cite{leewick, lee1, lee2, coleman}.
These authors used the term `acausal' in describing 
the behaviour.  We find this choice of terminology
unfortunate, since it may encourage 
 the incorrect conclusion that such theories might allow, for example,
for past events to be influenced by future actions.  Such effects 
are not possible.

As we will discuss in this section, it may be 
preferable to 
think of  theories of this type  not as acausal
but rather  as non-local.  Using Lee's original terminology, states may
contain initially undetectable `precursors' that are exponentially
amplified by time evolution and become relevant 
at later times.  To emphasize that causality is 
satisfied, we will perform a typical  thought experiment in 
a simple toy model and illustrate how 
acausal inconsistencies are avoided. 

For the reader uncomfortable with this aspect 
of these theories, we point out that  locally 
undetectable precursor
degress of freedom  have been invoked 
to explain aspects of the bulk-boundary correspondence 
in String Theory holography \cite{precursors1, precursors2}.   
We further point out that 
Horowitz and  Maldacena recently proposed resolving
the black hole information paradox via a final state 
boundary condition \cite{maldacena}. 
 Again, this is not so alien
in the context of Lee-Wick theories, since 
 the Lee-Wick real energy constraint 
can be reinterpreted as incorporating 
a final state boundary condition, 
namely, no blow-up at future infinity \cite{BG}.

We will discuss the issues in a simple toy model that captures the essential features
of the field theory scattering discussed in section \ref{scattering}.  Consider a model that has three states, denoted 
$$
  \left\{\ket{\textrm{incoming}}, \ket{\textrm{outgoing}}, \ket{\textrm{ghost}} \right\},
$$
with inner product
$$
  \eta \equiv \left( 
  \begin{array}{rrr}
   1 & 0 & 0 \\
  0 & 1 & 0 \\
  0 & 0 & -1
  \end{array}
\right)
$$
We declare the physical state at time $t = 0$
to be 
$$\ket{\textrm{incoming}}+ \ket{\textrm{outgoing}}+ \ket{\textrm{ghost}}$$
and introduce a scattering matrix between times $-T$ and $T$ as
$$
  S \equiv W_+^{-1}W_-
$$
where
$$
  W_- \equiv \left(
   \begin{array}{ccc}
     1 & 0 & 0 \\
   0 & \half\left({1\over \epsilon} + \epsilon\right) 
       &\half\left({1\over \epsilon} - \epsilon\right)  \\
   0 & \half\left({1\over \epsilon} - \epsilon\right) 
       &\half\left({1\over \epsilon} + \epsilon\right) 
\end{array}
   \right)
$$
and 
$$
  W_+^{-1} \equiv \left(
   \begin{array}{ccc}
   \half\left({1\over \epsilon} + \epsilon\right)  & 0
       &-\half\left({1\over \epsilon} - \epsilon\right)  \\
   0 & 1 & 0 \\
  - \half\left({1\over \epsilon} - \epsilon\right)  & 0 
       &\half\left({1\over \epsilon} + \epsilon\right)  
  \end{array}
   \right),
$$
where 
$$
  \epsilon \equiv e^{-\gamma T}, \qquad \gamma > 0.
$$
Applying $W_-^{-1}$ and $W_+$ respectively to the physical state at time
$t=0$, the physical state at time $t = -T$ is obtained as
$$\ket{\textrm{incoming}}+ \epsilon \left(\ket{\textrm{outgoing}}
   + \ket{\textrm{ghost}} \right) 
$$
and the physical state at time $t = T$ is obtained as
$$\ket{\textrm{outgoing}}+ \epsilon \left(\ket{\textrm{incoming}}
   + \ket{\textrm{ghost}} \right) .
$$
As in the field theory discussed in section \ref{scattering}, the in and 
out states contain an $\epsilon$-contribution that vanishes 
asymptotically and cannot be locally distinguished by 
observers in the sufficiently far past and future.  As we discussed, 
these corrections  grow 
exponentially with time and cannot be omitted, unlike in
the case of a positive-definite theory.

The above scattering matrix is meant to capture the essentials
of what happens in a local field theory with the set of 
processes indicated in the following set of diagrams:
\begin{equation}
\parbox{25mm}
{
\begin{fmfchar*}(25,25)
  \fmftop{b}
  \fmfbottom{t2,t1}
  \fmf{fermion}{t1,mid,t2}
  \fmf{dashes,tension=3}{b,mid}
  \fmfdot{mid}
\end{fmfchar*}
}
+
\parbox{25mm}
{
\begin{fmfchar*}(25,25)
  \fmfbottom{b}
  \fmftop{t2,t1}
  \fmf{fermion}{t1,mid,t2}
  \fmf{dashes,tension=3}{b,mid}
  \fmfdot{mid}
\end{fmfchar*}
}
+
\cdots
\label{causal}
\end{equation}
Two particles can annihilate to form a ghost, and a ghost can decay into 
a pair of particles.  For example, the first diagram represents the 
transition 
$$
  \ket{\textrm{incoming}} \to
   \ket{\textrm{ghost}} 
$$
via the action of $W_+^{-1}$ at some time $t > 0$.  
Components of the state that decay in the far past or future 
are indicated by dotted lines.  

Here the vertical axis is meant to connote time flow. 
Note  in particular
that the outgoing particles are created at time $t < 0$ \textit{before} the 
incoming particles annihilate at time $t > 0$.  This
seems, at first glance, to be a signal of acausality.
  However, essential to the resolution of this issue
is the observation that the outgoing particles are not
created from the incoming particles.  In particular, 
incoming and outgoing particles are always in different 
branches of the superposition.  Each individual branch is
causal.  

In particular, the interpretation of the scattering process is 
\textit{not} as in the following picture, suggested
by Coleman \cite{coleman}, 
who imagined an unstable intermediate particle moving
into the past.

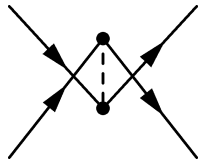
\begin{fmfchar*}(25,25)
  \fmftop{em,ec,ep}
  \fmf{fermion}{em,Zee,ep}
  \fmf{phantom,tension=4}{Zee,fc}
  \fmf{dashes}{Zee,Zff}
\fmf{phantom,tension=6}{Zff,ec}
  \fmf{fermion}{fb,Zff,f}
  \fmfbottom{fb,fc,f}
  \fmfdot{Zee,Zff}
\end{fmfchar*}

This diagram would suggest that there exists 
 an intermediate state
consisting of a tensor product of 
five particles, which would be incorrect.  
If one were to take such an interpretation seriously,  one would have to
explain 
what happens if an experimenter were to stop the incoming particles 
after the outgoing particles have already been emitted and observed.
Coleman suggested that the theory might 
incorporate  a novel uncertainty relation 
that would make such experiments impossible \cite{coleman}.  

Our analysis shows that
such an uncertainty relation is unnecessary.  In fact, it should be almost 
immediately clear that the version of the grandfather 
paradox described above is not possible, since 
at time $t = 0$ 
the incoming and outgoing scattered particles are in separate terms 
of a superposition.  Even if we could
observe the outgoing scattering products, we would not be 
able to then observe (and stop)
the incoming particles, since these are in separate branches.

In fact, it is impossible to observationally distinguish 
incoming 
from outgoing particles in the intermediate region.  The problem is that these 
individual 
terms are not physical states; indeed, the single
physical state 
at $t = 0$ is given by the combination $\ket{\textrm{incoming}}
+\ket{\textrm{outgoing}}+\ket{\textrm{null}}$.
To see what goes wrong if we were to try to 
distinguish the individual terms,  let us try to
couple the system to an observer who  measures, for example,
the outgoing particles at $t = 0$.  To see that such 
a measurement is unphysical, note that after 
the interaction, the 
combined state would have to be 
$$
  \ket{\textrm{incoming}}\otimes \ket{1}+\ket{\textrm{outgoing}}
\otimes \ket {0}+\ket{\textrm{null}}\otimes 
  \ket 0,
$$
where the second factor indicates the state of the observer.  
However, this combined state cannot be physical, since it diverges
exponentially 
under the subsequent evolution given by $W_+^{-1}\otimes \mathbf{1}$
as $T\to \infty$.  Unsurprisingly, consistency requires us to 
limit observations to physical states.  As a result, we cannot
locally measure outgoing and incoming states in the scattering region.

The question of locality should be addressed further.  Clearly, the physical states
are subject to constraints that may make them 
nonlocal, and they  cannot therefore be prepared and measured by 
entirely local experimenters.
For example, the $\epsilon$ term  of the initial state 
$$\ket{\textrm{incoming}}+ \epsilon \left(\ket{\textrm{outgoing}}
   + \ket{\textrm{ghost}} \right) 
$$
cannot be discarded without violating the physicality constraint,
leading to a divergent evolution
in the far future, but it is unclear how such a state could
be prepared by a pair of independent local agents, given
that the ghost is located at the center of 
mass of the two incoming particles.  However, we should 
remember that  experimenters  are subject to the 
same physicality constraints.  Thus,  the state of any two 
separated agents will itself contain a nonlocal term proportional to 
$\epsilon$ that emits the 
intermediate ghost.  This does not mean that these agents are not
independent.   Indeed, the $\epsilon$-term is uniquely
determined by the $\epsilon$-independent piece, and we could 
therefore label states in the far past by omitting the former.
  The resulting labeling will coincide with that of an ordinary
unconstrained field theory not containing ghosts, in which independent agents are 
usually assumed to be unproblematic. Given  such a labeling, however, 
the effect of the ghosts show up in the  effectively 
nonlocal interaction.

\section{Towards quantum black holes?}

The diagram that we have discussed has an interesting feature suggestive of
the expected behaviour of a microscopic black hole.  Of course, the theory
does not contain gravity, and the ghost particles do not have the correct quantum 
numbers or the ability to 
accrue matter as a black hole would.  However, the causal structure of the 
scattering diagram (\ref{causal})  
\begin{equation}
\parbox{25mm}
{
\begin{fmfchar*}(25,25)
  \fmftop{b}
  \fmfbottom{t2,t1}
  \fmf{fermion}{t1,mid,t2}
  \fmf{dashes,tension=3}{b,mid}
  \fmfdot{mid}
\end{fmfchar*}
}
+
\parbox{25mm}
{
\begin{fmfchar*}(25,25)
  \fmfbottom{b}
  \fmftop{t2,t1}
  \fmf{fermion}{t1,mid,t2}
  \fmf{dashes,tension=3}{b,mid}
  \fmfdot{mid}
\end{fmfchar*}
}
+
\cdots,
\label{blackhole}
\end{equation}
which is representative of, for example, the scattering of two 
high-energy leptons via their interaction with the complex-mass
 ghost 
positronium poles,
is reminiscent of that of a black hole being created and 
then evaporating.  In particular, the particles seen by the future observer
 are emitted from 
a point causally prior to the collision  of the incoming matter, which is reminiscent 
of  the way Hawking radiation originates causally prior to the singularity
that absorbs the incoming matter in a black hole.  Our model  even
incorporates an analogue of an evaporating  singularity, in the form  of the ghost
particle created by the collision of the incoming matter.  Indeed, 
we have seen that the amplitude of the ghost term  vanishes
in the far future, which is what one would expect from the 
singularity of an evaporating black hole.

Yet, the theory is by construction 
unitary with no loss of information or 
quantum coherence.  It is therefore perhaps possible to 
understand hypothetical 
features of quantum black holes with the help of relatively simple 
toy models based on these ideas.

Our observation shows some similarities with 
the work of Horowitz and  Maldacena \cite{maldacena}, 
on the black hole final state boundary condition.
Their
final 
state boundary condition ensured that everything 
falling into the black hole annihilated completely, 
leaving nothing behind.  
This is  indeed analogous  to what 
happens in the above diagrams,
where the incoming particles annihilate into a 
ghost whose amplitude decays exponentially 
in the future.

\section{Conclusion}

Given the results of this paper, we believe that there is
justification for reconsidering  quantum field theories
of  Lee-Wick type as potentially realistic  fundamental or effective 
theories of  the world.  
We believe that the ideas
described in this paper can be used to 
solve the previously open 
problem of quantizing and constraining
these theories non-perturbatively to obtain 
unitary theories that are also covariant. 

We illustrated the ideas by way of a Lee-Wick extension of QED,
which we argued to be asymptotically safe.
Basing the state space on certain  Gel'fand-Shilov
distributions, we showed explicitly that specific 
sets of ghost states can be covariantly eliminated
from the theory in a non-perturbatively well-defined way, 
and provided a plausible general argument, though not yet 
a rigorous proof, that all ghosts can indeed be eliminated 
this way, and that a unitary theory is obtained.  

We also described a  second, orthogonal   method for getting rid
of ghost states, based on a covariant maximum four-momentum 
constraint.  
In this approach, we pointed  out an interesting connection,
based on a naturalness  argument, between the largeness 
of the universe and the smallness of the fine structure
constant.  

We then discussed the so-called `acausality' of these
theories,  emphasizing that they are not really acausal, 
since
no inconsistencies can arise.  We propose that it may be 
less confusing to think of these theories as non-local at 
sufficiently high energy scales, with initial states 
containing precursors that may only become observable
at late times.  We pointed out some analogies between 
 Lee-Wick  precursors and the precursors invoked 
to explain aspects of the bulk-boundary correspondence 
in String Theory holography, and  between scattering 
processes in Lee-Wick theories mediated by ghosts 
and the behaviour of the black holes in  
Horowitz's and  Maldacena 's recently proposed 
resolution of the black
hole unitarity problem using final state 
boundary conditions.  
 As we pointed out, the causal 
structure of these kinds of Lee-Wick diagrams 
have aspects in common with the Penrose 
diagrams of black holes.  
Hopefully these comparisons will make the reader
less likely to reject Lee-Wick theories solely based on their
unusual causal behaviour.

Given our proposed resolutions to some prior 
foundational difficulties 
encountered in Lee-Wick theories, 
we believe that these theories deserve another look
in the search for descriptions of
nature.  Indeed, Lee-Wick theories  have recently enjoyed
a revival in the form of the 
Lee-Wick Standard Model
\cite{grinstein, espinosa, rizzo, dulaney, alvarez, carone, underwood, causality},
proposed to address, among other things,  the hierarchy problem and 
the stability  of the Higgs mass. 
Hopefully the results of this paper can be generalized to 
provide solutions to the problems of non-perturbative unitarity 
and Lorentz invariance in these theories.  

In the light of our results,  it may also be productive to revisit 
renormalizable
higher-derivative modifications of gravity, which 
contain Lee-Wick degrees of freedom
\cite{stelle}.

\section*{Acknowledgments}

    I would like to thank Antal Jevicki, Miquel Dorca, and the Brown
    University Physics department for their support.

\end{fmffile}

\begin{thebibliography}{99}


\bibitem{leewick}
T.D. Lee and G.C. Wick, \textit{Negative metric and the unitarity of the
S-matrix}, \textit{Nuclear Physics} \textbf{B9} (1969) 209-243.

\bibitem{lee1}
T.D. Lee, 
\textit{A finite theory of quantum electrodynamics},
in \textit{Elementary processes at high energy, part A,
  Ettore Majorana 1970 International School of Subnuclear Physics, Erice,
July 1-19}, Editor: A. Zichichi, Academic Press, New York and London, 1971.

\bibitem{lee2}
T.D. Lee, 
\textit{A relativistic complex pole model with indefinite metric},
in
\textit{Quanta, essays in theoretical physics dedicated 
to Gregor Wentzel},
P.G.O. Freund, C.J. Goebel, and Y. Nambu, eds.,
University of Chicago Press, 1970,
pp. 260-308.



\bibitem{PV}
   W. Pauli and F. Villars,
   \textit{On the invariant regularization in relativistic 
     quantum field theory},
   \textit{Reviews of Modern Physics} \textbf{21} (1949) 434-444.



\bibitem{cutkosky}
R.E. Cutkosky, P.V. Landshoff, D.I. Olive and J.C. Polkinghorne,
\textit{Nuclear Physics} \textbf{B12} (1969) 281-300.


\bibitem{BG}
D.G. Boulware and D.G. Gross, 
\textit{Lee-Wick indefinite metric quantization: A functional integral
approach},
\textit{Nuclear Physics} \textbf{B233} (1984) 1-23.

\bibitem{nakanishi}
N. Nakanishi, 
\textit{Lorentz noninvariance of the complex-ghost relativistic field theory},
\textit{Physical Review} \textbf{D3} (1971) 811-815.

\bibitem{nakanishiprime}
N. Nakanishi, 
\textit{Remarks on the complex-ghost relativistic field theory},
\textit{Physical Review} \textbf{D3} (1971) 3235-3237.

\bibitem{gleeson}
A.M. Gleeson, R.J. Moore, H. Rechtenberg, and E.C.G. Sudarshan, 
\textit{Analyticity, covariance, and unitarity in 
indefinite-metric quantum field theories},
\textit{Physical Review} \textbf{D4} (1971)  2242-2254.




\bibitem{gelfand1}
I.M Gel'fand and G.E. Shilov,
\textit{Generalized functions: Spaces of fundamental and generalized 
functions}, Vol. 2, Academic Press, Orlando, 1958.


\bibitem{gelfand2}
I.M Gel'fand and G.E. Shilov,
\textit{Generalized functions: Theory of differential equations}, Vol. 3, Academic Press, Orlando, 1967.


\bibitem{weinberg}
  S. Weinberg, in
  \textit{General Relativity},
S.W. Hawking and W. Israel, eds.
(Cambridge University Press, Cambridge. 1979),
p. 790.

\bibitem{coleman}
S. Coleman, 
\textit{Acausality}, in  
\textit{Subnuclear Phenomena, International School
of  Physics `Ettore Majorana' (1969; Erica, Italy}, 
New York, Academic Press, 1970,
pp. 282-327.

\bibitem{precursors1}
Joseph Polchinski, Leonard Susskind, Nicolaos Toumbas,
\textit{
Negative Energy, Superluminosity and Holography},
Phys.Rev. \textbf{D60} (1999) 084006.

\bibitem{precursors2}
Leonard Susskind, Nicolaos Toumbas,
\textit{Wilson Loops as Precursors},
Phys.Rev. \textbf{D61} (2000) 044001.

\bibitem{maldacena}
Gary T. Horowitz, Juan Maldacena,
\textit{The black hole final state},
JHEP \textbf{0402} (2004) 008.


\bibitem{bender1}
Carl M. Bender, Sebastian F. Brandt, Jun-Hua Chen, Qinghai Wang,
\textit{Physical Review} \textbf{D71} (2005) 025014.



\bibitem{grinstein}
B. Grinstein, D. O'Connell, M.B. Wise,
\textit{The Lee-Wick Standard Model},
Phys. Rev. \textbf{D77} (2008)  025012.

\bibitem{espinosa}
Jose Ramon Espinosa, Benjamin Grinstein, Donal O'Connell, Mark B. Wise,
\textit{Neutrino Masses in the Lee-Wick Standard Model},
Phys. Rev. D 77 (2008)  085002.

\bibitem{rizzo}
Thomas G. Rizzo,
\textit{Searching for Lee-Wick Gauge Bosons at the LHC},
arXiv:0704.3458v3 [hep-ph].

\bibitem{dulaney}
Timothy R. Dulaney, Mark B. Wise,
\textit{Flavor Changing Neutral Currents in the Lee-Wick Standard Model},
Phys. Lett. B 658 (2008)  230-235.


\bibitem{alvarez}
Ezequiel Alvarez, Leandro Da Rold, Carlos Schat, Alejandro Szynkman,
\textit{Electroweak precision constraints on the Lee-Wick Standard Model},
arXiv:0802.1061v1 [hep-ph].



\bibitem{carone}
Christopher D. Carone, Richard F. Lebed,
\textit {Minimal Lee-Wick Extension of the Standard Model},
arXiv:0806.4555v4 [hep-ph].




\bibitem{underwood}
T.E.J. Underwood and R. Zwicky,
\textit{Electroweak precision data and the Lee-Wick Standard Model},
arXiv:0805.3296v2 [hep-ph].


\bibitem{causality}
Benjamin Grinstein, Donal O'Connell, Mark B. Wise, 
\textit{ Causality as an emergent macroscopic phenomenon: The Lee-Wick O(N) model}
arXiv:0805.2156v1 [hep-th].


\bibitem{stelle}
K.S. Stelle,
\textit{Renormalization of higher-derivative quantum gravity},
Physical Review \textbf {D16} (1977)
953-969.

\bibitem{indef}
Andr\'e van Tonder and Miquel Dorca, \textit{Non-perturbative quantization of 
phantom and ghost theories: Relating definite and indefinite
      representations}, arXiv.org preprint hep-th/0610185.


\bibitem{bognar}
J. Bogn\'ar, \textit{Indefinite inner product spaces}, Springer-Verlag
Berlin, 1974.

\bibitem{malcev}
A.I. Mal'cev, \textit{Foundations of linear algebra}, 
W.H. Freeman and Company, 1963.

\bibitem{IZ}
C. Itzykson, J.-B. Zuber,
\textit{Quantum Field Theory},
International Edition 1985, McGraw-Hill (1980).









\bibitem{berggren}
 Tore Berggren, \textit{On the use of resonant states in eigenfunction 
expansions of scattering and reaction amplitudes},
\textit{Nuclear Physics} \textbf{A109} (1968) 265-287.

\bibitem{newton}
  Roger, G. Newton, \textit{Analytic properties of radial wave functions},
  \textit{Journal of Mathematical Physics} \textbf{1} (1960) 319-347 .

\bibitem{buchleitner}
A. Buchleitner, B. Gr\'emaud and D. Delande,
\textit{Wavefunctions of atomic resonances},
\textit{Journal of Physics} \textbf{B27} (1994) 2663-2679.


\bibitem{moiseyev}
Nimrod Moiseyev, 
\textit{Quantum theory of resonances: calculating energies, widths and
cross-sections by complex scaling},
\textit{Physics Reports} \textbf{302} (1998) 211-293.


\bibitem{hagen}
G. Hagen, J.S. Vaagen and M. Hjort-Jensen,
\textit{The contour deformation method in momentum space,
applied to subatomic physics},
\textit{Journal of Physics} \textbf{A37} (2004) 8991-9021.

\bibitem{ho}
Y.K. Ho, \textit{The method of complex coordinate rotation and its 
application to atomic collision processes},
\textit{Physics Reports} \textbf{99} (1983) 1-68.

\bibitem{madrid}
  {R. {de la Madrid}},
  \textit{Description of resonances within the rigged Hilbert space},
  arXiv.org preprint quant-ph/0607168,
  2006.

\bibitem{bogolubov}
  N.N. Bogolubov, A.A. Logunov, A.I. Oksak and I.T. Todorov,
\textit{General principles of quantum field theory},
English edition, Kluwer Academic Publishers, Dordrecht, 1990.



    \end{thebibliography}
\end{document}